%% file: main.tex
\documentclass[lettersize,journal]{IEEEtran}
\usepackage{amsmath,amsfonts}
\usepackage{algorithmic}
\usepackage{array}
\usepackage[caption=false,font=normalsize,labelfont=sf,textfont=sf]{subfig}
\usepackage{textcomp}
\usepackage{stfloats}
\usepackage{url}
\usepackage{verbatim}
\usepackage{graphicx}
\usepackage{multirow}
\usepackage{threeparttable}
\usepackage{multicol}
\usepackage{booktabs}
\usepackage{makecell}
\usepackage{orcidlink}
\usepackage{tablefootnote}
\usepackage{cite}
\usepackage{nicematrix,tikz}
\NiceMatrixOptions
  {
    custom-line = 
     {
       letter = : ,
       command = dashedline , 
       ccommand = cdashedline ,
       tikz = dashed
     }
  }

\def\BibTeX{{\rm B\kern-.05em{\sc i\kern-.025em b}\kern-.08em
    T\kern-.1667em\lower.7ex\hbox{E}\kern-.125emX}}

\begin{document}

\title{CleanMel: Mel-Spectrogram Enhancement for Improving Both Speech Quality and ASR}
\author{Nian~Shao,
Rui~Zhou,
Pengyu~Wang,
Xian~Li, 
Ying~Fang,
Yujie~Yang and 
Xiaofei~Li
\thanks{Nian Shao, Pengyu Wang, Ying Fang and Yujie Yang are with Zhejiang University, and also with Westlake University, Hangzhou, China (e-mail: \{shaonian, wangpengyu, fangying, yangyujie\}@westlake.edu.cn).}
\thanks{Rui Zhou, Xian Li and Xiaofei Li are with the School of Engineering, Westlake University, and also with the Institute of Advanced Technology, Westlake Institute for Advanced Study, Hangzhou, China (e-mail: \{zhourui, lixian, lixiaofei\}@westlake.edu.cn). }
\thanks{Nian Shao and Rui Zhou equally contributed to this work. Corresponding: lixiaofei@westlake.edu.cn}
}

\markboth{Journal of \LaTeX\ Class Files,~Vol.~18, No.~9, September~2020}%
{How to Use the IEEEtran \LaTeX \ Templates}

\maketitle

\begin{abstract}
In this work, we propose CleanMel, a single-channel Mel-spectrogram denoising and dereverberation network for improving both speech quality and automatic speech recognition (ASR) performance. The proposed network takes as input the noisy and reverberant microphone recording and predicts the corresponding clean Mel-spectrogram. The enhanced Mel-spectrogram can be either transformed to the speech waveform with a neural vocoder or directly used for ASR. The proposed network is composed of interleaved cross-band and narrow-band processing in the Mel-frequency domain, for learning the full-band spectral pattern and the narrow-band properties of signals, respectively. Compared to linear-frequency domain or time-domain speech enhancement, the key advantage of Mel-spectrogram enhancement is that Mel-frequency presents speech in a more compact way and thus is easier to learn, which will benefit both speech quality and ASR. Experimental results on five English and one Chinese datasets demonstrate a significant improvement in both speech quality and ASR performance achieved by the proposed model.
Code and audio examples of our model are available online \footnote{https://audio.westlake.edu.cn/Research/CleanMel.html}.
\end{abstract}

\setcounter{table}{0}
\input{FinalFiles/1_introduction}

\input{FinalFiles/2_problem_formulation}

\input{FinalFiles/3_method}

\input{FinalFiles/4_experiment}

\input{FinalFiles/5_results}

\input{FinalFiles/6_conclusion}

% \clearpage
% Bibliography
\bibliographystyle{IEEEtran}
\bibliography{refs}

\end{document}

%% file: FinalFiles/1_introduction.tex
\section{Introduction}

\IEEEPARstart{T}HIS work studies single-channel speech enhancement using deep neural networks (DNNs) to improve both speech quality and automatic speech recognition (ASR) performance. 
A large class of speech enhancement methods employ DNNs to map from noisy and reverberant speech to corresponding clean speech, conducted either in the time domain \cite{Luo2019Conv, defossez20demucs} or time-frequency domain \cite{Xiong2022Spectro, hu2020DCCRN, li2022glance}. While these methods can efficiently suppress noise, they do not necessarily improve ASR performance due to the speech artifacts/distortions introduced by speech enhancement networks \cite{iwamoto22analysis}. In \cite{kinoshita2020improving}, it was found that time-domain enhancement is more ASR-friendly than frequency-domain enhancement. A time-domain progressive learning method is proposed in \cite{nian2022progressive}, which also shows the superiority of time-domain speech enhancement, and the progressive learning mechanism is very effective for robust ASR by mitigating the over-suppression of speech. In \cite{yang2023robustASR, yang2024towards}, ASR performance is largely improved by decoupling front-end enhancement and backend recognition. In \cite{yang2024towards}, it is shown that an advanced time-frequency domain network, i.e., CrossNet, can even outperform the time-domain attentive recurrent network (ARN) \cite{yang2023robustASR}.

Speech enhancement in Mel-frequency domain, or similarly in
rectangular bandwidth (ERB) domain, has been developed under various contexts in the literature. Mel-frequency and ERB bands model human speech perception of spectral envelope and signal periodicity, within which speech enhancement is more perceptually and computationally efficient than within linear-frequency domain or time domain. In \cite{valin2018hybrid,valin2020perceptually, schroter23deepfilter}, spectral envelope enhancement is performed within the ERB bands, and then applying pitch filtering \cite{valin2018hybrid, valin2020perceptually} or deep filtering \cite{schroter23deepfilter} to recover the enhanced speech. 

Sub-band networks \cite{Xiong2022Spectro,li2019narrow} and full-band/sub-band fusion networks, i.e., FullSubNet \cite{hao2021fullsubnet, zhou2023rts}, have been proposed and achieved outstanding performance. However, separately processing sub-bands in the linear-frequency domain leads to a large computational complexity. To reduce the number of sub-bands and thus the computational complexity, Fast FullSubNet \cite{hao2022fast} and the work of \cite{kothapally2023deep} proposed to perform sub-band processing in the Mel-frequency domain, and then transform back to linear frequency with a joint post-processing network. In \cite{liu22voicefixer, tian2023diffusion}, speech enhancement is directly conducted in the Mel-frequency domain and then a separately trained neural vocoder is used to recover speech waveform. These methods improve speech enhancement capability by alleviating the burden of enhancing full-band speech details and also by leveraging the powerful full-band speech recovery capacity of advanced neural vocoder, as a result, achieve higher speech quality compared to their linear-frequency counterparts. 

In this work, we propose CleanMel, a single-channel Mel-spectrogram enhancement network for improving both speech quality and ASR performance. Different from the previous works \cite{valin2018hybrid,valin2020perceptually,schroter23deepfilter,hao2022fast} that performing speech enhancement in the ERB or Mel-frequency domain, and then applying a joint pitch filtering or deep filtering to obtain the enhanced speech, this work decouples the Mel-spectrogram enhancement and post-processing parts by targeting the enhancement network with clean Mel-spectrogram. The enhanced Mel-spectrogram can be directly used for ASR, or transformed back to waveform with a separate neural vocoder as is done in \cite{liu22voicefixer}. Compared to linear-frequency spectrogram or time-domain waveform, Mel-frequency spectrogram presents speech in a more compact and less-detailed way (but still perceptually efficient) and has a lower feature dimension (number of frequencies) from the perspective of machine learning, which would result in lower prediction error. 
This is beneficial for both speech quality improvement and ASR: (i) Neural vocoders have been extensively studied in the field of Text-to-Speech, and are capable of transforming Mel-spectrogram back to time-domain waveform with sufficient speech details and naturalness. Therefore, the low-error property of Mel-spectrogram enhancement can be hopefully maintained by the neural vocoders and higher speech quality can be achieved. (ii) From the perspective of ASR, there seems no need to first recover the less-accurate full-band (or time-domain sample-wise) speech details and then compress to Mel-frequency. A direct and more-accurate Mel-spectrogram estimation would be preferred.
 
We adapt the network architecture of our previous proposed (online) SpatialNet \cite{quan2024spatialnet,quan2024oSpatialNet} with some modifications to better accommodate Mel-spectrogram enhancement.
SpatialNet is composed of interleaved cross-band and narrow-band blocks originally proposed for processing multichannel STFT frequencies. The narrow-band block processes STFT frequencies independently to learn the spatial information presented in narrow-band (one frequency), such as the convolutive signal propagation and the spatial correlation of noise. And the cross-band block is designed for learning the across-frequency dependencies of narrow-band information. 
As for single-channel Mel-spectrogram enhancement in this work, the narrow-band block processes Mel frequencies independently to also learn the (single-channel) convolutive signal propagation of target speech, which is crucial not only for conducting dereverberation of target speech but also for discriminating between target speech and interfering signals. The cross-band block is now reinforced and utilized to learn the full-band spectral pattern in the Mel-frequency domain.  

Overall, in this work, we propose a single-channel Mel-spectrogram enhancement network, with an advanced network architecture adapted from SpatialNet \cite{quan2024spatialnet,quan2024oSpatialNet}. The enhanced Mel-spectrogram can be directly fed into a pre-trained ASR model, or transformed to waveform via a neural vocoder. In addition, we have studied several critical issues when decoupling the Mel-spectrogram enhancement front-end and ASR/Vocoder back-ends. i) We have systematically studied and compared different learning targets that can be used for Mel-spectrogram enhancement, including logMel mapping, Mel ratio masking and the clipping issue of logMel; ii) We have developed a data normalization scheme to align the signal levels of cascaded front-end and back-end models; iii) We have developed an online neural vocoder to enable online speech enhancement.

Experiments are conducted on six public datasets (five English and one Chinese) for speech denoising and dereverberation individually or jointly. Importantly, we adopt a more realistic evaluation setup: from multiple data sources of clean speech, real-measured room impulse responses (RIRs) and noise signals, we collect and organize a relatively large-scale training set, based on which we train the network for once and directly test it on all the six test sets. Experiments show that the proposed model achieves the state-of-the-art (SOTA) speech enhancement performance in term of speech perceptual quality. 
Moreover, on top of various pre-trained and advanced ASR models, the proposed model prominently improves the ASR performance on all datasets. These results demonstrate that our trained models have the potential to be directly employed to real applications. 

%% file: FinalFiles/2_problem_formulation.tex
\section{Problem formulation}

The noisy and reverberant single-channel speech signals can be represented in the time domain as
\begin{align}
\label{eq:td}
y(n) = s(n) * a(n) + e(n)
\end{align}
where $n$ stands for the discrete time index. $s(n)$ and $e(n)$ represents the clean source speech and ambient noise, respectively. $a(n)$ denotes the RIR and $*$ the convolution operation. In this work, only static speaker is considered, thence the RIR is time-invariant. 
RIR is composed of the direct-path propagation, early reflections and late reverberation.

We conduct joint speech denoising and dereverberation in this work, which amounts to estimate the (Mel-spectrogram of) desired direct-path speech $x(n)=s(n) * a_\text{dp}(n)$ from microphone recording $y(n)$, where $a_\text{dp}(n)$ denotes the direct-path part in RIR. The training target for both the proposed network and neural vocoders are derived with $x(n)$.  

The proposed method is performed in the time-frequency domain. By applying STFT to Eq.~(\ref{eq:td}), based on the convolutive transfer function approximation \cite{li2019multichannel}, we can obtain:
\begin{align}
\label{eq:stft}
Y(f,t) \approx S(f,t) * A(f,t) + E(f,t)
\end{align}
where $f\in\{0,...,F-1\}$ and $t\in\{1,...,T\}$ denote the indices of frequency and time frame, respectively. $Y(f,t)$, $S(f,t)$ and $E(f,t)$ are the STFT of respective signals, and $X(f,t)$ is the STFT of direct-path speech.  $A(f,t)$ is the convolutive transfer function associated to $a(n)$. Convolution $*$ is conducted along time. In the STFT domain, the time-domain convolution $s(n) * a(n)$ is approximately decomposed as frequency-wise convolutions $S(f,t) * A(f,t)$. We consider one STFT frequency as one narrow-band, and refer to the frequency-wise convolution as narrow-band convolution.
Speech dereverberation in this work highly relies on learning this narrow-band convolution.  
For noise reduction, one important way for discriminating between speech and stationary noise is to test the signal stationarity, which can be modeled in narrow-band as well.

%% file: FinalFiles/3_method.tex
\section{Mel-spectrogram Enhancement}
In this work, we propose to enhance the Mel-spectrograms, which then can be directly fed into an ASR model, or transformed to waveforms with a neural vocoder. 

\subsection{Learning Target: Clean Mel-spectrogram}
\label{sec: learning_target}
The power-based or magnitude-based Mel-spectrogram of the target speech $X(f,t)$, denoted as $X_{\text{mel}}(f,t)$, can be obtained by weighted summing the squared magnitude $|X(f,t)|^2$ or magnitude $|X(f,t)|$ over frequencies with the triangle weight functions of Mel filterbanks,  where $f\in\{1,...,F_{\text{mel}}\}$ now indexes  Mel frequency. Our preliminary experiments showed that using power-based or magnitude-based Mel-spectrograms achieve similar enhancement performance, which means we can choose either of them according to the Mel setup of ASR or neural vocoder backend models. In this work, we use the power-based Mel-spectrogram according to the setup of commonly-used ASR models. Then, the logMel-spectrogram, namely the logarithm of $X_{\text{mel}}(f,t)$ 
\begin{align}
\label{eq:logmel}
X_{\text{logmel}}(f,t)=\log(\max\{X_{\text{mel}}(f,t), \epsilon\})
\end{align}
can be taken as the input feature of ASR or neural Vocoder backends. The base of logarithm ($e$ or 10) should be consistent to the one of back-ends as well, while $e$ is used in this work. 
The Mel-spectrogram is clipped with a small value of $\epsilon$ to avoid applying logarithm to close-zero values. 

Normally, $\epsilon$ is set to be very small, e.g. 1e-10, to maintain complete speech information. However, in our preliminary experiments, we found that very small speech values are not very informative for both ASR and neural vocoder, and can be clipped without harming performance. Moreover, since those small values are highly contaminated by noise or reverberation, the prediction error of them could be very large. 
For these reasons, we set $\epsilon$ to a relatively large value, e.g. 1e-5 (when the maximum value of time-domain signal is normalized. The signal normalization methods will be presented in Section~\ref{sec:normalization}). Fig.~\ref{fig: cleantarget} gives an example of our target logMel-spectrogram, in which about 40\% TF bins are clipped. 

In this work, we evaluate two different learning targets as for Mel-spectrogram enhancement. 

\textbf{LogMel mapping}: The clean logMel-spectrogram can be directly predicted with the network. The training loss is set to the mean absolute error (MAE) loss between the predicted and the clean logMel-spectrogram, namely 
\begin{align}
\begin{aligned}
    \mathcal{L}_{\text{MAE}}
    =\frac{1}{F_{\text{mel}}T}\sum_{f=1}^{F_\text{mel}}\sum_{t=1}^T  |\hat{X}_{\text{logmel}}(f,t) - X_{\text{logmel}}(f,t)|,     
\end{aligned}
\label{eq: loss}
\end{align}
where $\hat{X}_{\text{logmel}}(f,t)$ is the network output. 

\textbf{Mel ratio mask}: Ratio mask is one type of popular learning targets for speech magnitude enhancement  \cite{wang2018supervised}. For each time-Mel-frequency bin, the Mel ratio mask is defined as 
\begin{align}
    M(f, t) = \text{min} \left( \sqrt{\frac{X_{\text{mel}}(f,t)}{Y_\text{mel}(f, t)}}, 1 \right),
\end{align}
where $Y_\text{mel}(f, t)$ denotes the power level of noisy Mel-spectrogram. The square root function transforms the power domain to the magnitude domain. The $\text{min}(\cdot)$ function rectifies the mask into the range of $[0,1]$. 
The mean squared error (MSE) of the ratio mask is taken as the training loss, namely
\begin{align}
    \mathcal{L}_{\text{MRM}} = \frac{1}{F_{\text{mel}}T}\sum_{f=1}^{F_\text{mel}}\sum_{t=1}^T (M(f, t)-\hat{M}(f, t))^2,
\end{align}
where $\hat{M}(f, t)$ denotes the model prediction of ${M}(f, t)$. Then, the enhanced logMel-spectrogram can be obtained as
\begin{align}
    \hat{X}_\text{logmel}(f, t) = \text{log}(\max\{\hat{M}(f, t)^2Y_\text{mel}(f, t), \epsilon\}).
\end{align}

\input{FinalFiles/Figs/M_logmel_spec}

\subsection{The CleanMel Network} 
\label{sec: model_arch}
\input{FinalFiles/Figs/M_model_architecture}

Fig.~\ref{fig: model_arch} shows the network architecture. The proposed network takes (the real ($\mathcal{R}(\cdot)$) and imaginary ($\mathcal{I}(\cdot)$) parts of) the STFT of microphone recording as input, i.e. $Y(f,t)$, denoted as $\mathbf{y}$:
\begin{align}
\label{eq:feature}
\mathbf{y}[f,t,:]=[\mathcal{R}(Y(f,t)), \mathcal{I}(Y(f,t))]\in \mathbb{R}^2,
\end{align}
where $[:]$ represents to take values of one dimension of a tensor.
The network is composed of an input layer, interleaved cross-band and narrow-band blocks first in the linear-frequency domain and then in Mel-frequency domain, a Mel-filterbank, and a Linear output layer. The input layer performs temporal convolution on $\textbf{y}$ with kernel size 5, producing the hidden representation with the dimensions of $F\times T \times H$.
Then one cross-band block and one narrow-band block process the hidden tensor in the linear-frequency domain. A Mel-filterbank (with triangle weight functions) transforms the frequency dimension from $F$ linear frequencies to $F_\text{mel}$ Mel frequencies via non-trainable matrix multiplication. Then, $L$ interleaved cross-band and narrow-band blocks process the tensor in Mel-frequency domain.

After the final narrow-band block, the output Linear layer transforms the $H$-dimensional features to 1-dimensional output, producing either the enhanced logMel-spectrogram or the Mel ratio mask. Note that a sigmoid activation is applied when predicting the Mel ratio mask. 

\subsubsection {Narrow-band block}
\label{sec: arch_narrow_band}
As shown in Eq.~(\ref{eq:stft}), the time-domain convolution can be decomposed as frequency-independent narrow-band convolutions, which have much lower complexity than time-domain convolution in terms of room filter order. Therefore, modeling narrow-band convolution is substantially more efficient than modeling time-domain convolution. The convolution model of target speech not only provides necessary information for dereverberation but also helps discriminate target speech from other interfering sources. Additionally, in narrow-band processing, non-stationary speech and stationary noise can be well discriminated by testing the signal stationarity. For these reasons, we propose the narrow-band network, which processes frequencies independently along the time dimension, with all frequencies sharing the same network. 

The narrow-band convolution in Eq.~(\ref{eq:stft}) is defined between the complex-valued STFT coefficients of source speech and room filters. Therefore, we process the complex-valued STFT coefficients of noisy signal (in a hidden space) rather than other features such as magnitude, to preserve the convolution property. We first process in the finer linear-frequency domain using one narrow-band block to fully exploit the convolution information. After Mel-filterbank transformation, we process in the coarser Mel-frequency domain with additional narrow-band blocks.

The narrow-band network comprises Mamba blocks \cite{gu2023mamba}. Mamba is a recently proposed network architecture based on structured state space sequence models, which has proven highly efficient for learning both short-term and long-term dependencies in sequential data. 
Besides short-term correlations of signals, there exist some long-term dependencies should be exploited for narrow-band speech enhancement. For example, the narrow-band convolution remains time-invariant over the entire recording for static speakers. Moreover, Mamba has a linear computational complexity w.r.t time, which is suitable for streaming processing of long audio signals. 
Specifically, each narrow-band block contains a forward Mamba for online processing and an optional backward Mamba (output averaged with forward output) for offline processing.

\subsubsection{Cross-band block}
\label{sec: arch_cross_band}
The cross-band block learns full-band/cross-band signal dependencies. The original SpatialNet \cite{quan2024spatialnet} designed this block to learn the linear relationship of inter-channel features (e.g. inter-channel phase difference) across frequencies. In this work, there is no such inter-channel information for the single-channel scenario. Instead, the cross-band block can learn full-band spectral patterns in linear/Mel-frequency domains, which is also critical for (especially single-channel) speech enhancement. The cross-band block processes frames independently along the frequency dimension, with all frames sharing the same network.

Specifically, we retain the cross-band block architecture of SpatialNet, which comprises cascaded frequency convolutional layers (F-GConv1d), across-frequency linear layers (F-Linear), and a second frequency convolutional layer. The frequency convolutional layers perform 1-D convolution along frequency to capture adjacent-frequency correlations, while the across-frequency linear layer processes all frequencies jointly (for each hidden dimension separately) to model full-band dependencies.

One cross-band block is first applied in the linear-frequency domain to learn detailed full-band correlations. To reduce the model complexity, the hidden dimension $H$ is compressed (e.g., to $H/12$) before processing by the across-frequency linear layer ($F^2$ complexity for each hidden dimension).  
After Mel-filterbank transformation, cross-band blocks learn full-band correlations across Mel frequencies, reducing the complexity from $F^2$ to $F_\text{mel}^2$. Correspondingly, we set the hidden dimension to remain $H$ (without compression) to enhance the full-band learning capability.  

All the Mel-frequency cross-band blocks share identical across-frequency linear layers.

\subsection{Back-ends}
\label{sec: model_backends}

At inference, CleanMel is followed by either an ASR model or a neural vocoder. Both ASR model and neural vocoder are trained separately from the CleanMel network. 

\subsubsection{ASR}
During inference, the enhanced logMel-spectrogram can be directly fed to a pre-trained ASR model, without fine-tuning or joint training.
Different ASR systems may use varying STFT configurations, numbers of Mel frequencies, and logarithm bases. To seamlessly integrate the enhanced logMel-spectrogram into one ASR model, CleanMel adopts identical configurations to the target ASR model.   
Training CleanMel is not very costly, so it can be easily re-trained for a new ASR system, especially for those large-scale or deployed ASR systems.
In this work, we only conduct offline ASR combined with offline CleanMel.

\subsubsection{Neural vocoder}
\label{sec: model_vocos}
The vocoder we adopt in this work is Vocos \cite{siuzdak2024vocos}, a recently proposed  Generative Adversarial Network (GAN)-based neural vocoder. The generator of Vocos predicts the STFT coefficients of speech at frame level and then generates waveform through inverse STFT, thence it significantly improves the computational efficiency compared to HiFi-GAN that directly generates waveform at sample level. Vocos uses the multiple-discriminators and multiple-losses proposed in HiFi-GAN \cite{kong2020hifi}. In this work, to unify the front-end and back-end processing, we have made several necessary modifications to Vocos as follows:
\begin{itemize}
\item The magnitude-based Mel-spectrogram of original Vocos is modified as power-based to match front-end and ASR configurations, where the two cases were shown to achieve similar performance in our preliminary experiments. 
\item The sampling rate of signals, the STFT configurations and the number of Mel-frequencies of Vocos are adjusted to match the front-end and ASR configurations. 
\item The original Vocos is designed for offline processing, as it employs non-causal convolution layers. To enable online processing, we modified Vocos to be causal by substituting each non-causal convolution layer with its causal version. The modified online vocoder still performs well. In addition, to reduce the computational complexity of online processing, the 75\% STFT overlap of original Vocos is reduced to be 50\% overlap, which still achieves comparable performance.
\end{itemize}
The offline and online Vocos are used to work with the offline and online CleanMel, respectively. All Vocos models were trained using our direct-path target speech $x(n)$.

\subsection{Signal Normalization}
\label{sec:normalization}
When cascading separately-trained front-end and back-end models, signal normalization should be performed not only to facilitate the training of respective models but also to align the signal level of cascaded models.
Specifically, the level of enhanced logMel of CleanMel should align with the one of the input logMel of ASR and Vocoder models.
 For offline processing, CleanMel adopts the normalization method of Vocos: a random gain is applied to time-domain signal such that the maximum level of the resulting signal falls between -1 and -6 dBFS (decibels relative to full scale). This ensures the peak sample value approaches but remains less than 1, so that the generated waveform can be directly played at full volume without clipping. CleanMel applies this normalization to the noisy signal, and uses the same gain of noisy signal to the  corresponding clean target signal. In this way, the enhanced logMel can be directly fed to Vocos. When applying this time-domain normalization, the logMel clip value $\epsilon$ is set to 1e-5 for both CleanMel and Vocos. ASR models normally have a separate logMel normalization, which will be used to re-normalize the enhanced logMel. 

For online processing, the time-domain normalization is no longer applicable. Instead, an online STFT-domain normalization is used for both CleanMel and Vocos. For CleanMel, the noisy and target speech are normalized in the STFT domain as $\tilde{Y}(f,t)=Y(f,t) / \mu(t)$ and $\tilde{X}(f,t)=X(f,t) / \mu(t)$, where $\mu(t)$ is recursively updated as the mean of STFT magnitude, i.e., $\mu(t)=\alpha \mu(t-1)+(1- \alpha) \frac{1}{F} \sum_{f=0}^{F-1} |Y(f,t)|$. The smoothing weight is set to $\alpha=\frac{K-1} {K+1}$ which is equivalent to applying a $K$-long rectangle smoothing window. When training online Vocos with target speech $x(n)$, we first apply the offline time-domain normalization mentioned above, followed by this online normalization. Specifically, $\mu(t)$ is computed with and applied to $X(f,t)$ (instead of $Y(f,t)$), and then the corresponding logMel is computed as the input of Vocos generator. 
Accordingly, the output of Vocos generator (before applying inverse STFT) would be an estimation of normalized $X(f,t)$. To go back to the signal level of time domain normalization, the recursive normalization factor $\mu(t)$ is multiplied back to the estimation of normalized $X(f,t)$ and then applying inverse STFT, after which Vocos losses (including Mel loss and discriminator losses) are computed. During inference, online normalization is applied to noisy inputs of CleanMel. The enhanced logMel is directly fed into Vocos.
The recursive normalization factor computed with the noisy input is multiplied to the estimated STFT coefficients by Vocos and then applying inverse STFT to obtain the final waveform. 
This final waveform is time-domain-normalized and ready for playback. For online normalization, the logMel clip value $\epsilon$ is set to $1e-4$ for both CleanMel and the input of Vocos generator.

%% file: FinalFiles/Figs/M_logmel_spec.tex
\begin{figure}[t]
\begin{center}
    \includegraphics[width=\linewidth]{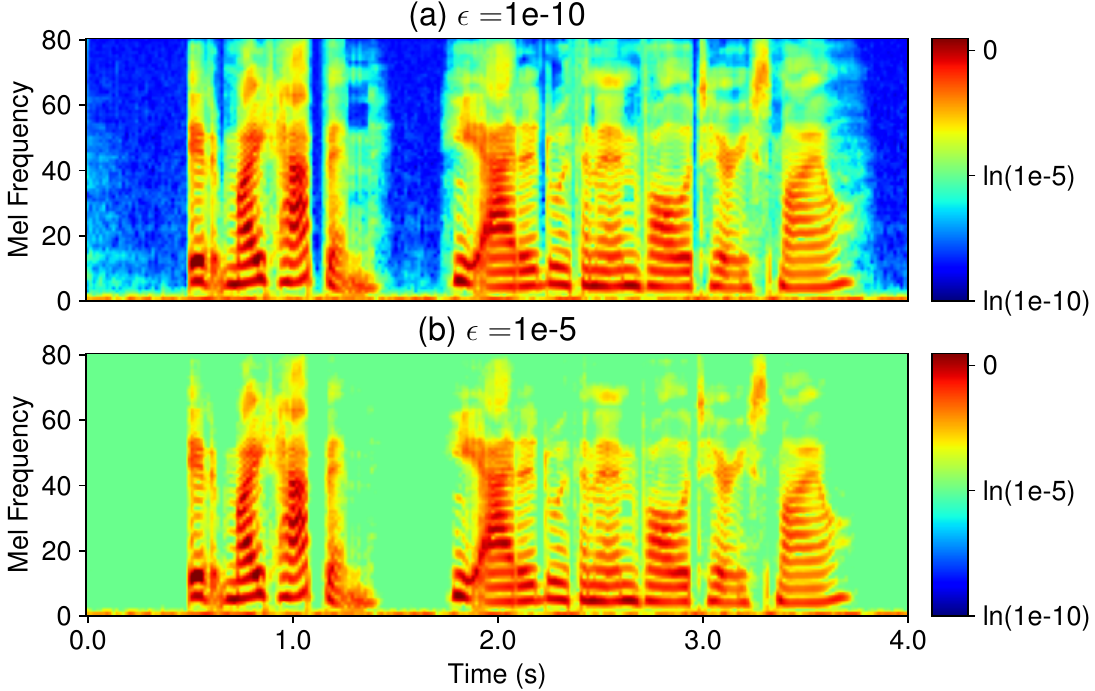}
    \vspace{-2em}
    \caption{An example of logMel spectrogram with a clip value of $\epsilon=$1e-10 or $\epsilon=$1e-5.}
    \label{fig: cleantarget}
\end{center}
\end{figure}

%% file: FinalFiles/Figs/M_model_architecture.tex
\begin{figure*}[t]
\begin{center}
    \includegraphics[width=0.8\linewidth]{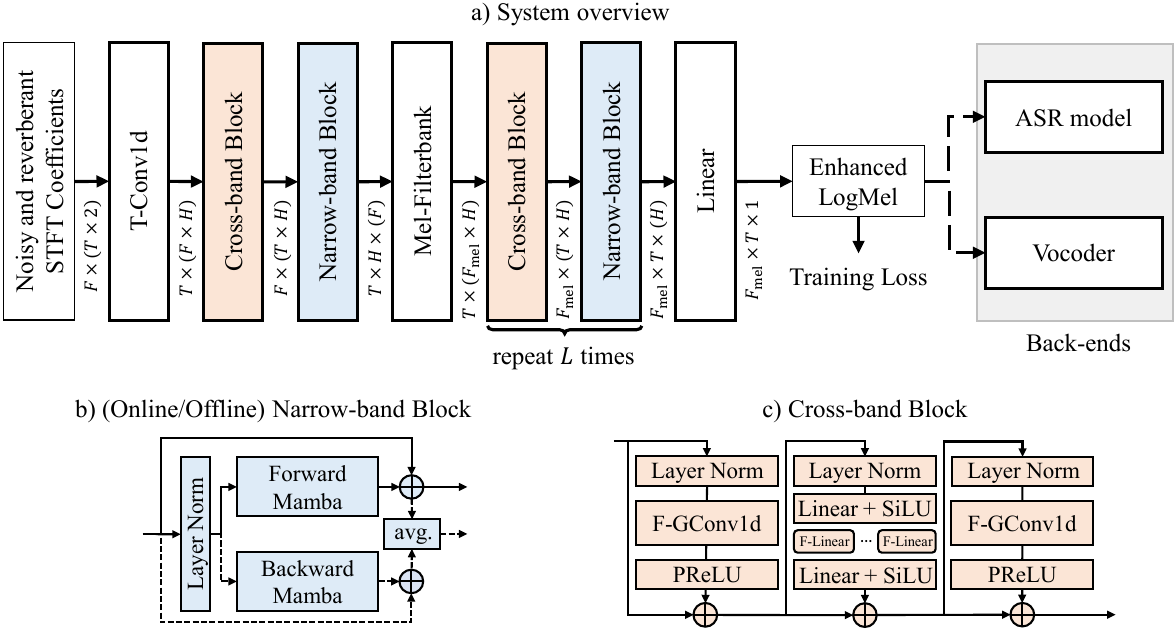}
    
    \caption{Model architecture of the proposed speech enhancement system. a) System overview. The input dimension of neural blocks/layers are presented before each of them in the form `batch dimension x (dimension of one sample in batch)". b) The online/offline narrow-band block. The solid lines stands for online processing. The solid plus dashed lines stands for offline processing. c) The cross-band block. }
    
    \label{fig: model_arch}
\end{center}
\end{figure*}

%% file: FinalFiles/4_experiment.tex
\section{Experimental setups}
In the section, we present the experiment datasets, experimental configurations, evaluation metrics and comparison methods. 

\subsection{Dataset}

\subsubsection{Speech enhancement training dataset} The proposed model is trained with synthetic noisy/clean speech pairs. Reverberant speech signals are generated by convolving source speech signals with RIRs, then added with noise signals. Clean target speech signals are generated by convolving source speech signals with the direct-path part of RIRs.

In this work, we conduct speech enhancement for both Mandarin Chinese and English. Six different datasets (as will be shown later) are used to evaluate our model. We attempt to train the model once and test it on all the datasets, which will reflect the general capability of the model under various situations. To do this, we collect data (in terms of source speech, RIRs and noise) from multiple public datasets, and form a training set with sufficient speech quality and environment/device diversity.

\textbf{Source speech}: Source speech signals are collected from 6 datasets, including AISHELL I \cite{bu2017aishell1}, AISHELL II \cite{du2017aishell2} and THCHS30 \cite{wang2015thchs} for Chinese, and EARS \cite{richter2024ears}, VoiceBank \cite{voicebank} and DNS I challenge \cite{reddy2020dns1} for English. 

For each language, about 200 hours of high-quality speech data are selected from the original datasets, based on the raw score of DNSMOS P.835 \cite{reddy2022dnsmos835}. The selection thresholds are set to 3.6 and 3.5 for Chinese and English data, respectively. Except that the entire EARS training set is included, since EARS involves various emotional speech that cannot be well evaluated by the DNSMOS. The two test speakers in the VoiceBank Demand dataset \cite{vctk-demand} are excluded from our training data.

The amount of data selected from each dataset is summarized in Table~\ref{tab: e_speech_dur}.

\input{FinalFiles/Tables/E_speech_dur_table}

\textbf{RIR}: 
We use real-measured RIRs from multiple public datasets \cite{eaton2015ace, jeub2009air, prawda2022ARNI, carlo2021dechorate, Hadad2014MultiChannel, james2016naturalreverb, Kinoshita2013REVERB, nakamura2000rwcp}. Table~\ref{tab: e_rirs} shows the statistics of these RIR datasets. For all (even multi-channel) datasets, all RIRs are used, except that we uniformly sampled 1,000 RIRs from the very large original ARNI dataset. The reverberation time, i.e. $T_{60}$, of RIRs mostly lie in the range between 0 and 1.5 seconds, except for a few rooms in AIR \cite{jeub2009air} and NaturalReverb \cite{james2016naturalreverb}. The distribution of the $T_{60}$s are shown in Fig.~\ref{fig: t60_dist}. 
Besides the wide distribution range of $T_{60}$s, these RIRs also have large diversity in terms of environments and measuring devices. For example, NaturalReverb \cite{james2016naturalreverb} is recorded in 271 spaces encountered by humans during daily life, AIR \cite{jeub2009air} is measured with a dummy head for binaural applications, RWCP \cite{nakamura2000rwcp} use a Head-Torso as source speaker, etc. 
When synthesizing reverberant speech, 80\% source speech samples are convolved with a randomly selected RIR, while there is no RIR convolution for the rest 20\% samples to account for the near-field applications where reverberation is negligible. 
\input{FinalFiles/Tables/E_rirs}
\input{FinalFiles/Figs/E_rir_dist}

\textbf{Noise}:
Speech and noise are mixed at a random signal-to-noise ratio (SNR) between -5 dB and 20 dB. We use the noise signals from the DNS challenge \cite{reddy2021dns3} and the RealMAN dataset \cite{Bing2024RealMAN}. The DNS challenge dataset has about 181 hours of noise sampled from AudioSet \cite{audioset} and Freesound. The RealMAN dataset has 106 hours of ambient noise recorded in 31  daily life scenes, including various indoor, semi-outdoor, outdoor and transportation scenes.  

\subsubsection{Speech enhancement evaluation datasets}
The speech enhancement performance of the proposed model is evaluated on six datasets. For Chinese, we use the public test set (static speaker) of the RealMAN dataset \cite{Bing2024RealMAN}. For English, the evaluation is conducted on 5 different datasets: (1) the CHiME4 challenge `isolated\_1ch\_track' test set \cite{vincent2016chime4}, (2) the one-channel test set of REVERB \cite{Kinoshita2013REVERB}, (3) the test set of DNS I challenge \cite{reddy2020dns1}, (4) the public EARS blind test dataset \cite{richter2024ears}, (5) the test set of VoiceBank Demand dataset \cite{vctk-demand}.

\subsubsection{ASR evaluation datasets and models}
The ASR performance of the proposed model is evaluated on three datasets. For Chinese, we conduct experiments on the RealMAN (static speaker) dataset  \cite{Bing2024RealMAN}, using the pre-trained WenetSpeech Conformer ASR model \cite{Zhang2022wenet} from ESPNet \footnote{https://github.com/espnet/espnet/tree/master/egs2}. For English, we evaluate on the CHiME4 \cite{vincent2016chime4} and REVERB \cite{Kinoshita2013REVERB} datasets, using their respective pre-trained E-branchformer \cite{kim2023branchformer} and Transformer \cite{vaswani2017attention} ASR models from ESPNet, respectively. Note that, the ASR models used in this work have the same STFT and Mel-frequency configurations, thus one CleanMel model can be used for all of them.

\subsection{Configurations}

\input{FinalFiles/Tables/E_model_size}

\subsubsection{Data configurations}
\label{sec:dataset_data_preprocessing}
The sampling rate of all data is set to 16 kHz. STFT is applied using Hanning window with a length of 512 samples (32 ms) and a hop size of 128 and 256 samples (8 and 16 ms) for the offline and online models, respectively. The offline model has a finer temporal resolution than the online model since it is used for ASR in this work and its temporal resolution is aligned with the ASR models. However, we empirically found that, compared to the 16-ms hop size, the 8-ms hop size does not benefit much for the speech enhancement performance. The number of Mel frequencies is set to $F_\text{mel}=80$ for the frequency range of 0-8 kHz. The same STFT implementation (ESPNet implementation\footnote{https://github.com/espnet/espnet/blob/master/espnet2/layers/stft.py})
is used for the CleanMel networks, neural vocoders and ASR models, to avoid configuration mismatch. The natural logarithm (base of $e$) is used.

\subsubsection{Network configurations}
Follow \cite{quan2024spatialnet}, the kernel size of the T-Conv1d in the input module and the F-GConv1d layers in the cross-band block are both set to 5. As shown in Fig.~\ref{fig: model_arch}, in narrow-band block, forward-only and forward/backward Mamba layers are set for online and offline processing, respectively. We set up two model scales for the offline models, referred to as CleanMel-{S} and CleanMel-{L}. The online model scale is set approximately to CleanMel-{S}. The configurations are shown in Table~\ref{tab: model_size}. The depth $L$ of online models are set to twice the one of corresponding offline models to have the similar model size. Due to the different setups of STFT hop size, the computational complexity, i.e. FLOPs, of offline models are roughly twice the one of corresponding online models.

\subsubsection{Training and inference setups}
For CleanMel, AdamW optimizer \cite{loshchilov2018AdamW} with an initial learning rate of $10^{-3}$ is used for training. The learning rate exponentially decays with $\text{lr} \leftarrow 0.001 \times 0.99^{\text{epoch}}$. Gradient clipping is applied with a gradient norm threshold 10. The batch size are set to 32. Training samples are synthesized in an on-the-fly manner, and 100,000 samples are considered as one training  epoch. The CleanMel-{S} and CleanMel-{L} models are trained by 100 and 150 epochs, respectively. 
Afterward, we average the model weights of the last 10 epochs as the final model for inference. 

For training the Vocos neural vocoder \cite{siuzdak2024vocos}, we synthesized 400,000 (direct-path) clean speech samples with both English and Chinese data used for CleanMel training. The training configurations remain unchanged as in its original work. 

\subsection{Evaluation metrics}
Speech enhancement performance is evaluated with Perceptual Evaluation of Speech Quality (PESQ) \cite{rix2001pesq}, DNSMOS P.808 \cite{reddy2021dnsmos808} and P.835 \cite{reddy2022dnsmos835}, where the background (BAK), signal (SIG) and overall (OVL) scores for P.835 are all reported. Word Error Rate (WER) and Character Error Rate (CER) are used to evaluate English and Chinese ASR performances, respectively.

\subsection{Comparison models}
We compare with advanced speech enhancement models, most of which were claimed in their original papers to be able to conduct joint speech denoising and dereverberation, including 
(i) ConvTasNet \cite{Luo2019Conv} is a fully convolutional time-domain (offline) speech separation network.
(ii) FullSubNet \cite{hao2021fullsubnet} is a LSTM-based full-band and sub-band fusion network originally proposed for online speech denoising, and extended to speech dereverberation in \cite{zhou2023rts}. For offline processing, we change the uni-directional LSTMs to be bi-directional. 
(iii) Demucs \cite{defossez20demucs} is an online time-domain speech enhancement network with a U-net architecture.
(iv) CMGAN \cite{CMGAN} is a two-stage Conformer-based offline speech enhancement network.
(v) LiSenNet \cite{yan2025lisennet} is a super light-weight online speech enhancement network primirily with a dual-path recurrent module. 
Both CMGAN and LiSenNet utilize mixed training losses, including speech regression losses and an adversarial loss. 
(vi) VoiceFixer \cite{liu22voicefixer} also enhances Mel-spectrograms (using a ResUNet) and generates the waveform using a neural vocoder, but supports only offline processing. 
(vii) CDiffuse \cite{cdiffuse}, StoRM \cite{Lemercier2023storm} and UNIVERSE++ \cite{universe++} are three diffusion-based offline speech enhancement models.
(viii) SpatialNet \cite{quan2024spatialnet} and oSpatialNet \cite{quan2024oSpatialNet} perform speech enhancement in the STFT linear-frequency domain, and offer the backbone network for the proposed CleanMel model. In SpatialNet and oSpatialNet, self-attention (and temporal convolution) and Mamba are adopted for learning the narrow-band spatial information in an offline and an online manner, respectively. 

For offline processing, besides comparing with the original SpatialNet, we also implemented a variant of it with bi-directional Mamba for narrow-band blocks (same architecture with the proposed method), which is referred to as SpatialNet-Mamba,  and serves as the linear-frequency baseline for the proposed Mel-frequency model.
Note that, SpatialNet and oSpatialNet were originally proposed for multi-channel speech enhancement, and this work is the first one to fully evaluate and analyze the capability of them for single-channel speech enhancement. 

To conduct fair comparisons, we re-trained the ConvTasNet, FullSubNet, CMGAN, LiSenNet, CDiffuse, StoRM, UNIVERSE++ and oSpatialNet/SpatialNet models, using the same training data utilized for the proposed model. LiSenNet was trained without the noise detector \cite{yan2025lisennet}. UNIVERSE++ was trained under its universal speech enhancement configuration \cite{universe++}. All models were trained to convergence according to their model selection strategy.
However, we found that it is not easy to re-train the Demucs and Voicefixer models. In \cite{defossez20demucs}, different data augmentation strategies are applied when training Demucs on different datasets. Voicefixer \cite{liu22voicefixer} was designed for 44.1 kHz signals. Re-training Demucs and VoiceFixer requires careful data engineering or hyperparameter search. Therefore, we used the pre-trained checkpoints of Demucs and VoiceFixer to perform speech enhancement on only English data. For Demucs, we used the `dns64' model provided by the authors \footnote{https://github.com/facebookresearch/denoiser}. For VoiceFixer, we empolyed the default pre-trained model provided in the open-source package \footnote{https://github.com/haoheliu/voicefixer}. Following the VoiceFixer default inference setup, test waveforms were first up-sampled to 44.1 kHz to perform speech enhancement and then down-sampled back to 16 kHz for evaluation.

%% file: FinalFiles/Tables/E_speech_dur_table.tex
\begin{table}[t]
\setlength\tabcolsep{2.5pt}
    \centering
    \caption{Clean source speech used in this work.}
    \vspace{-1em}
    \label{tab: e_speech_dur}
    \begin{tabular}{c|c|c|c}
    \toprule
     Language & Speech dataset & Duration (hours) & \#Speakers  \\
     \midrule
     \multirow{3}{*}{Chinese} & AISHELLI \cite{bu2017aishell1} & 25 & 240 \\
     & AISHELL II \cite{du2017aishell2} & 178 & 987 \\
     & THCHS30 \cite{wang2015thchs} & 18 & 49 \\
     \midrule
     \multirow{3}{*}{English} & EARS \cite{richter2024ears} & 81 & 92\\
     & VoiceBank \cite{voicebank} & 32 & 93 \\
     & DNS I challenge \cite{reddy2020dns1} & 94 & 1,263 \\
     \bottomrule
    \end{tabular}

\end{table}

%% file: FinalFiles/Tables/E_rirs.tex
\begin{table}[t]
    \centering
    \caption{Real-measured RIRs used in this work. }
    \vspace{-1em}
    \label{tab: e_rirs}
    \begin{tabular}{c|c|c}
    \toprule
     RIR dataset & \#RIRs & $T_{60}$ range (second) \\
     \midrule
 ACE \cite{eaton2015ace} & 84 & $[0.33, 1.22]$\\
 AIR \cite{jeub2009air}  & 204 & $[0.08, 4.50]$\\
 ARNI \cite{prawda2022ARNI} & 1,000 & $[0.51, 1.20]$ \\
 dEchorate \cite{carlo2021dechorate} & 594 & $[0.17, 0.75]$ \\
 MultiChannel \cite{Hadad2014MultiChannel} & 234 & ${0.16, 0.36, 0.61}$ \\ 
 NaturalReverb \cite{james2016naturalreverb} & 270 & [0.02, 2.00] \\
 REVERB \cite{Kinoshita2013REVERB} & 240 & [0.25, 0.70] \\
 RWCP \cite{nakamura2000rwcp} & 182 & [0.10, 0.72] \\
     \bottomrule
    \end{tabular}

\end{table}

%% file: FinalFiles/Figs/E_rir_dist.tex
\begin{figure}[t]
\begin{center}
    \includegraphics[width=0.8\linewidth]{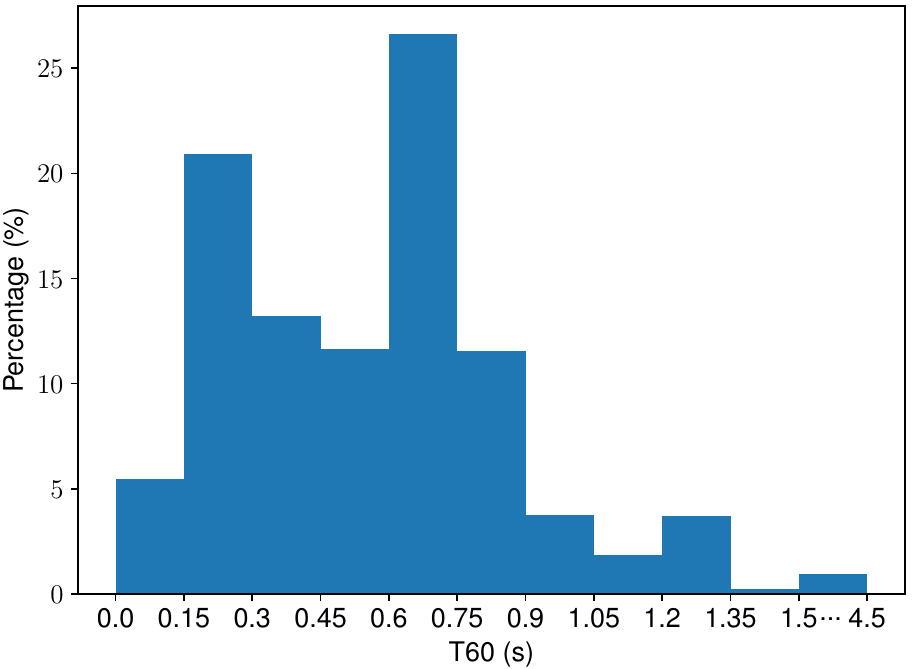}
    \caption{T60 distribution of RIRs used in this work.}
    \label{fig: t60_dist}
\end{center}
\end{figure}

%% file: FinalFiles/Tables/E_model_size.tex
\begin{table}[t]
    \centering
    \label{tab: model_size}
    \caption{Model configurations of the proposed Mel spectrogram enhancement networks.}
    \vspace{-1em}
    \setlength\tabcolsep{2.5pt}
    \begin{tabular}{c|c|cccc}
        \toprule 
        Mode & Model size & Depth($L+1$) & Hidden($H$) & \#Param(M) & FLOPs(G/s) \\
        \midrule
       Online
       & CleanMel-S & 16  & 96  & 2.7  & 18.1  \\
        \midrule
       \multirow{2}{*}{Offline} 
       & CleanMel-S & 8  & 96   & 2.5  & 32.9 \\
       & CleanMel-L & 16 & 144  & 7.2  & 127.8 \\
       \bottomrule
    \end{tabular}
\end{table}

%% file: FinalFiles/5_results.tex
\section{Results and Analyses}

In this section, we present and analyze speech enhancement performance and ASR results, conduct ablation studies, and compare the model size and computational complexity.

\subsection{Speech enhancement results}
\input{FinalFiles/Tables/R_SE_DNSMOS}

\input{FinalFiles/Tables/R_SE_PESQ}
Table~\ref{tab: se_mos} and \ref{tab: se_pesq} show the DNSMOS and PESQ scores, respectively. We analyze the results in the following aspects:
\subsubsection{Comparing the training targets of logMel mapping and Mel ratio mask} For online processing, \emph{mapping} consistently outperforms \emph{mask} in DNSMOS, mainly due to the higher residual noise of \emph{mask}, which can be reflected by the much lower BAK scores of \emph{mask} on the very noisy CHiME4 and RealMAN test sets. When applying the predicted ratio mask on the noisy spectrogram, if speech is highly contaminated by noise, there will exist certain residual noise even if the predicted error of mask is small. By contrast, \emph{mapping} directly predicts the logMel of speech, which can avoid such residual noise. However, as shown in Table~\ref{tab: se_pesq}, \emph{mask} achieves higher PESQ scores than \emph{mapping} on the highly noisy CHiME4 and DNS (w.o. reverb) sets. PESQ measures the perceptual similarity of enhanced speech and reference clean speech. The higher PESQ scores indicates that \emph{mask} performs better on retrieving the target speech through extracting the target speech from the noisy speech with a mask. Directly \emph{mapping} may erroneously remove or boost those speech components highly contaminated by noise. The enhanced logMel-spectrograms are transformed to waveforms with a neural vocoder, and the speech quality measured with DNSMOS is more affected by the residual noise caused by \emph{mask} than the \emph{mapping} error of logMel-spectrogram. Fig.~\ref{fig: r_mask_mapping} shows an example of the enhanced logMel-spectrogram by \emph{mask} and \emph{mapping}, which verifies our discussions. 

\input{FinalFiles/Figs/R_mask_mapping} 

For offline processing, the differences between \emph{mask} and \emph{mapping} discussed above still present, but get much smaller, for example the difference of BAK scores on the CHiME4 and RealMAN test sets become much smaller. Compared to the online network, the offline network achieves much smaller prediction errors for both \emph{mask} and \emph{mapping}, thence their differences also become smaller as they both target the same clean speech.

\subsubsection{Comparing with oSpatialNet and SpatialNet-Mamba} oSpatialNet and SpatialNet-Mamba adopt the same network architectures and serve as the linear-frequency baselines for the proposed CleanMel networks. For both online and offline processing, the proposed Mel-spectrogram enhancement networks noticeably outperform their linear-frequency baselines in DNSMOS, but not necessarily in PESQ. This phenomenon is consistent with the findings in the VoiceFixer \cite{liu22voicefixer}. The GAN-based nerual vocoder is good at generating high-quality speech, but will possibly reduce the  similarity of the enhance speech with the reference speech (PESQ). 

\subsubsection{Comparing with online baseline networks}  Demucs \cite{defossez20demucs} performs well for denoising but not for dereverberation, as it achieves leading performance in DNSMOS P.835 on the (low-reverberation) CHiME4, EARS and VoiceBank Demand sets, but not for other reverberant test sets. As a light-weight network, LiSenNet \cite{yan2025lisennet} achieves promising results. On several test sets, the performance measures of LiSenNet approach even surpass the ones of FullSubNet \cite{hao2021fullsubnet}. Although it was developed for multi-channel speech enhancement in \cite{quan2024oSpatialNet}, oSpatialNet also performs very well for single-channel speech enhancement compared with other comparison models. oSpatialNet and the proposed CleanMel network work better especially for dereverberation due to their narrow-band processing of room filter convolution.

\subsubsection{Comparing with offline baseline networks} Among the three discriminative methods (ConvTasNet \cite{Luo2019Conv}, FullSubNet \cite{hao2021fullsubnet} and CMGAN \cite{CMGAN}), the more recently proposed CMGAN performs the best and achieves top-tier performance measures.  
VoiceFixer \cite{liu22voicefixer} is an important baseline for the proposed CleanMel network, as they both perform Mel-spectrogram enhancement. VoiceFixer adopts an advanced ResUNet. We can see that the proposed network noticeably outperforms VoiceFixer, which verifies the strong capability of the proposed cross-band and narrow-band combination network.
Among the three diffusion-model-based generative methods (CDiffuse \cite{cdiffuse}, StoRM \cite{Lemercier2023storm} and UNIVERSE++ \cite{universe++}), StoRM and UNIVERSE++ perform better than CDiffuse. The proposed network consistently outperforms StoRM. Their performance gaps of DNSMOS scores are not large, but the gaps of PESQ scores are very large. This indicates that although generative model can generate high-quality speech, but its fidelity to the source speech is hard to be ensured.
SpatialNet \cite{quan2024spatialnet}, especially its Mamba variant, i.e. SpatialNet-Mamba, as a pure discriminative network, achieves promising results. 

Overall, by combining a powerful discriminative Mel-spectrogram enhancement network and a GAN-based generative neural vocoder, the proposed network achieves new SOTA speech enhancement performance for both online and offline processing, and for both denoising and dereverberation. By scaling up the proposed network, the performance can be further improved. Compared to the small models, the large models noticeably improve the PESQ scores (although not the DNSMOS scores).    

\input{FinalFiles/Tables/R_ASR}

\subsection{ASR results}
\label{sec: r_asr}
Table~\ref{tab: asr_all} presents the ASR results. We can observe that the ASR results are consistent with the common finding in the field \cite{iwamoto22analysis} that single-channel speech enhancement networks usually do not help for ASR due to the speech artifacts caused by the networks. VoiceFixer performs Mel-spectrogram enhancement using an advanced ResUNet. The poor ASR results of VoiceFixer show that ResUNet brings severe artifacts to the enhanced Mel-spectrogram. StoRM does not work well for ASR, which further verifies that the generated speech has low fidelity to the source speech. 

Three exceptions are CMGAN, SpatialNet and SpatialNet-Mamba. They show clear positive ASR effects on REVERB-real. SpatialNet and SpatialNet-Mamba also show positive ASR effects on RealMAN. We believe that the advantages of CMGAN and SpatialNet lie in their narrow-band processing mechanism (the Conformer-Time block in CMGAN and the narrow-band block in SpatialNet): i) narrow-band processing is especially efficient for modeling the narrow-band room filter convolution; ii) as discussed in \cite{li2019narrow}, narrow-band processing can avoid the so-called wide-band artifacts, such as the blurred and/or wrongly deleted/inserted wide-band spectra, which will be very harmful for ASR.  

The proposed CleanMel networks further enhance ASR performance beyond SpatialNet-Mamba. Directly enhancing the Mel-frequency spectra is more optimal for ASR than first enhancing the detailed linear-frequency spectra and then compressing to Mel-frequency. 
Compared to linear-frequency processing, the lower-dimensional Mel-frequency representation is easier to learn. However, this is more valid for the cross-band processing part, but not for the narrow-band (frequency-wise or dimension-wise) processing part. 
That is possibly why the proposed CleanMel networks are more advantageous on the denoising task of CHiME4 (relying more on cross-band processing) than the dereverberation task of REVERB (relying more on narrow-band processing). For ASR, Mel ratio \emph{mask} performs better than logMel \emph{mapping}, but the performance gaps are not large. This indicates that ASR is less affected by the larger residual noise caused by \emph{mask} than the prediction error of logMel \emph{mapping}. Scaling up is very effective for further improving the ASR performance. 

Overall, by adopting an advanced backbone network (i.e. SpatialNet-Mamba), developing an effective Mel-spectrogram enhancement framework, and scaling up the model size, the proposed CleanMel-L-mask network achieves substantial ASR improvements relative to unprocessed speech.

\input{FinalFiles/Figs/R_mel_linear_loss}

\subsection{Comparing linear- and Mel-frequency speech enhancement}
We have mentioned that, compared to linear-frequency enhancement, Mel-frequency enhancement is easier to learn due to its lower frequency dimension, which is especially valid for the cross-band learning module in the proposed model. 
To verify this, we compare the validation loss (prediction error) of conducting enhancement in 80, 160 Mel-frequency domain and 256 linear-frequency domain. The CleanMel-S-mask model is adopted. Since the masking ratio (defined in Eq.~(\ref{eq: loss})) has a constant range of values, i.e. $[0,1]$, its prediction error in different frequency domains can be directly compared.  

\indent The validation losses for the first 100 K training steps are shown in Fig.~\ref{fig: mrm_compare}. It can be seen that the prediction errors are indeed lowered as the number of frequencies is decreased. Note that, the computational complexity of enhancement models are roughly proportional to the number of frequencies as well.  

\subsection{Influence of logMel clip value}
\input{FinalFiles/Tables/R_ASR_clip_value}
As described in Sec.~\ref{sec: learning_target}, we employ a larger clip value ($\epsilon=1e-5$) when computing logMel compared to a regular setup ($\epsilon=1e-10$). 
This section evaluates the influence of this setting.  Experiments are conducted on the clean source speech used in CHiME4 test set (i.e. WSJ0 \cite{chime_clean} speech) and REVERB simulated test set (i.e. WSJCAMO \cite{reverb_clean} speech). Clean logMel provides the upperbound performance for logMel enhancement, thence the influence of the clip value to clean logMel reflects its potential influence to enhanced logMel. Specifically, clean logMels with different clip values ($1e-5$ and $1e-10$) are fed to neural vocoders and ASR models, to obtain speech waveforms and ASR results, respectively. Note that two neural vocoders are respectively trained for the two clip values. The pre-trained ASR models (with clip value of $1e-10$) in CHiME4 and REVERB datasets are directly used for both clip values. Results are presented in Table~\ref{tab: asr_clip_value}. It can be seen that the two clip values have negligible performance differences, which verifies our statement that a larger clip value of $1e-5$ does not degrade the speech quality (through a neural vocoder) and ASR performance, since the very small values (smaller than $1e-5$) are not very informative. In addition, trying to enhance the very small values (more contaminated by noise and reverberation) increases the task difficulty.

\subsection{Influence of neural vocoder to ASR performance}
\input{FinalFiles/Tables/R_ASR_wav}
We use a GAN-based neural vocoder to transform the enhanced Mel-spectrogram to waveform, which may modify the original Mel-spectrogram and reduce the fidelity, although a Mel-spectrogram loss is adopted in the vocoder to maintain the Mel-spectrogram fidelity. Table~\ref{tab: asr_wav} compares the ASR results between taking as input the enhanced Mel-spectrogram and the vocoder waveform. Note that the ASR models used for VoiceFixer is re-trained according to its STFT and Mel-frequency setups. We can see that the neural vocoders indeed reduce the Mel-spectrogram fidelity as well as the ASR performance. However, the performance reductions are not significant, and the ASR results of vocoder waveform for each model closely align with the results of its corresponding enhanced Mel-spectrogram. This means the vocoder waveform could be also an alternative/suboptimal choice for ASR. If it is not easy to re-train a CleanMel model according to the setups of a new ASR model, since vocoder waveform is irrelevant to those setups, our pre-trained checkpoints can be used.  

\input{FinalFiles/Tables/R_ASR_finetune}
\input{FinalFiles/Tables/R_model_complexity}

\subsection{Joint fine-tuning of the CleanMel and ASR models}
As shown in the previous works \cite{chang2022end,shi2024waveform}, joint fine-tuning of the cascaded speech enhancement and ASR models can mitigate speech distortions introduced by the speech enhancement model, and further improve the ASR performance. In this section, we investigate the fine-tuning strategies of the proposed model. Experiments are conducted on the CHiME4 dataset, with the pretrained CleanMel-S-mask model, and the pretrained E-branchformer ASR model of CHiME4. \\
\indent Three strategies are evaluated: 1) freezing the ASR model and fine-tuning CleanMel-S-mask, 2) freezing CleanMel-S-mask and fine-tuning the ASR model and 3) joint fine-tuning of the CleanMel and ASR models. Fine-tuning is implemented in the ESPNet framework. We replace the STFT and logMel transformations of the original ASR system by our pretrained CleanMel model. Data augmentation of SpecAug \cite{park19specaug} is applied to the enhanced logMel, but time-warping is not applied. All models are trained with the AdamW \cite{loshchilov2018AdamW} optimizer. 
We observed that faster convergence is achieved when the ASR model is unfrozen. 
Accordingly, ASR-only fine-tuning and joint fine-tuning last for 2 epochs and CleanMel-only fine-tuning lasts for 4 epochs. For all experiments, the learning rate ramps up from 0 to 1e-5 during the first epoch (2,300 steps), then decays according to the principle set in ESPNet. Models are validated on the development set for every 230 steps. The best 10 checkpoints are averaged as the final model. \\
\indent Results are presented in Table~\ref{tab: asr_ft}. It can be seen that ASR performance can be further improved by all strategies. 
This demonstrates that, although the pretrained CleanMel model can largely improve ASR performance over the unprocessed signal, the enhanced logMel still contains certain artifacts detrimental to ASR. These artifacts can be further mitigated through fine-tuning. Among the three strategies, only fine-tuning the CleanMel model achieves the best performance. However, this may not be conclusive, different results could be obtained for other ASR models or datasets.

\subsection{Model size and computational complexity} 
Table~\ref{tab: model_complexity} shows the model size and computational complexity of the proposed model and comparison models. The proposed model is composed of a Mel-spectrogram enhancement network and an optional neural vocoder. The Mel-spectrogram enhancement network has a small model size and a large computational complexity, mainly due to the independent computation of frequencies with shared narrow-band blocks. Compared to linear-frequency sub-band/narrow-band processing networks, including FullSubNet, oSpatialNet, SpatialNet(-Mamba), the proposed Mel-spectrogram enhancement network has a smaller computational complexity, due to the much less frequencies to be processed, i.e. 80 Mel frequencies versus 256 linear frequencies. As for diffusion models, the computation cost is much more expensive than other models.

%% file: FinalFiles/Tables/R_SE_DNSMOS.tex
\begin{table*}[t]
\begin{center}
\setlength\tabcolsep{1pt}
\renewcommand\arraystretch{1.2}
\caption{{DNSMOS scores on real and simulated test sets.}}
\vspace{-1em}
\label{tab: se_mos}
\begin{tabular}{c|c|cccc|cccc|cccc|cccc|cccc|cccc}

\toprule
& 
\multirow{4}{*}{\makecell[t]{\textbf{Enhancement} \\ \textbf{method}}} &
\multicolumn{4}{c|}{\multirow{2}{*}{CHiME4 real}} &
\multicolumn{4}{c|}{\multirow{2}{*}{EARS}} &
\multicolumn{4}{c|}{\multirow{2}{*}{\makecell{{VoiceBank} \\ {Demand}}}} &
\multicolumn{4}{c|}{\multirow{2}{*}{REVERB real}} &
\multicolumn{4}{c|}{\multirow{2}{*}{DNS w. rvb.}} & 
\multicolumn{4}{c}{\multirow{2}{*}{RealMAN static}} \\

& & & & & & & & & & & & & & & & & & & & & & & & & \\

\cline{3-26}
&
& \multicolumn{3}{c}{P.835} & \multicolumn{1}{|c|}{\multirow{2}{*}{P.808}}
& \multicolumn{3}{c}{P.835} & \multicolumn{1}{|c|}{\multirow{2}{*}{P.808}}
& \multicolumn{3}{c}{P.835} & \multicolumn{1}{|c|}{\multirow{2}{*}{P.808}}
& \multicolumn{3}{c}{P.835} & \multicolumn{1}{|c|}{\multirow{2}{*}{P.808}}
& \multicolumn{3}{c}{P.835} & \multicolumn{1}{|c|}{\multirow{2}{*}{P.808}}
& \multicolumn{3}{c}{P.835} & \multicolumn{1}{|c }{\multirow{2}{*}{P.808}} \\
\cline{3-5} \cline{7-9} \cline{11-13} \cline{15-17} \cline{19-21} \cline{23-25}
&
& OVL & SIG & BAK & \multicolumn{1}{|c|}{}
& OVL & SIG & BAK & \multicolumn{1}{|c|}{}
& OVL & SIG & BAK & \multicolumn{1}{|c|}{}
& OVL & SIG & BAK & \multicolumn{1}{|c|}{}
& OVL & SIG & BAK & \multicolumn{1}{|c|}{}
& OVL & SIG & BAK & \multicolumn{1}{|c}{}  \\

\midrule
& unproc.
& 1.37 & 1.99 & 1.34 & 2.43
& 2.02 & 2.94 & 2.04 & 2.80
& {2.53} & {3.28} & {2.87} & {3.04}
& 1.39 & 1.74 & 1.53 & 2.81
& 1.42 & 1.89 & 1.50 & 2.74
& 1.53 & 2.03 & 1.75 & 2.56 \\

\midrule

\multirow{6}{*}{\rotatebox{90}{\textbf{Online}}} & FullSubNet \cite{hao2021fullsubnet}
& 2.41 & 3.05 & 3.08 & 3.22
& 2.73 & 3.12 & 3.55 & 3.52
& {2.93} & {3.30} & {3.79} & {3.47}
& 2.72 & 3.13 & 3.60 & 3.61
& 2.41 & 2.98 & 3.15 & 3.36
& 2.49 & 2.97 & 3.33 & 3.12 \\

& Demucs \cite{defossez20demucs}
& 2.84 & 3.12 & \textbf{3.90} & 3.26
& \textbf{2.94} & 3.21 & \textbf{3.94} & 3.38
& {\textbf{3.03}} & {\textbf{3.31}} & {3.97} & {3.43}
& 2.96 & 3.25 & 3.94 & 3.50
& 2.50 & 2.82 & 3.66 & 3.17
& \multicolumn{4}{c}{-} \\

& {LiSenNet \cite{yan2025lisennet}}
& {2.26} & {2.70} & {3.31} & {3.08}
& {2.61} & {2.93} & {3.63} & {3.29}
& {2.87} & {3.16} & {3.93} & {3.42}
& {2.54} & {3.15} & {3.31} & {3.38}
& {2.21} & {2.56} & {3.49} & {3.15}
& {2.32} & {2.70} & {3.42} & {2.96} \\

& oSpatialNet \cite{quan2024oSpatialNet}
& 2.61 & 3.15 & 3.36 & 3.52
& 2.85 & 3.22 & 3.65 & 3.65
& {2.91} & {3.26} & {3.80} & {3.48}
& 3.03 & 3.33 & 3.93 & 3.84
& 2.79 & 3.16 & 3.70 & 3.67
& 2.79 & 3.19 & 3.66 & 3.46\\

\cline{2-26}

& CleanMel-S-map
& \textbf{2.95} & \textbf{3.26} & 3.86 & \textbf{3.77}
& 2.89 & \textbf{3.24} & 3.69 & \textbf{3.72}
& {3.02} & {3.28} & {\textbf{4.01}} & {\textbf{3.65}}
& \textbf{3.17} & \textbf{3.45} & \textbf{4.02} & \textbf{3.94}
& \textbf{3.08} & \textbf{3.36} & \textbf{3.97} & \textbf{3.84}
& \textbf{2.90} & \textbf{3.24} & \textbf{3.81} & \textbf{3.52} \\
% & 2.92 & 3.23 & 3.86 & 3.75
% & 2.89 & 3.24 & 3.69 & 3.71
% & 3.14 & 3.41 & 4.02 & 3.92
% & 3.02 & 3.31 & 3.94 & 3.81
% & 2.87 & 3.21 & 3.81 & 3.49 \\

& CleanMel-S-mask
& 2.74 & 3.26 & 3.43 & 3.59
& 2.83 & 3.19 & 3.64 & 3.70
& {2.99} & {3.31} & {3.90} & {3.50}
& 3.07 & 3.40 & 3.91 & 3.80
& 2.95 & 3.30 & 3.81 & 3.75
& 2.73 & 3.18 & 3.53 & 3.36 \\

\midrule
\midrule

\multirow{13}{*}{\rotatebox{90}{\textbf{Offline}}}

& {ConvTasNet \cite{Luo2019Conv}}
& {2.79} & {3.19} & {3.67} & {3.12}
& {2.81} & {3.19} & {3.62} & {3.30}
& {3.00} & {3.31} & {3.93} & {3.38}
& {2.95} & {3.25} & {3.92} & {3.29}
& {2.50} & {2.90} & {3.51} & {3.05}
& {2.64} & {3.06} & {3.54} & {2.97} \\

& FullSubNet \cite{hao2021fullsubnet}
& 2.47 & 3.10 & 3.11 & 3.27
& 2.74 & 3.16 & 3.51 & 3.54
& {3.03} & {3.34} & {3.93} & {3.52}
& 2.69 & 3.09 & 3.60 & 3.60
& 2.35 & 2.91 & 3.10 & 3.25
& 2.47 & 2.97 & 3.27 & 3.13 \\

& {CMGAN \cite{CMGAN}}
& {3.11} & {3.48} & {3.82} & {3.55}
& {\textbf{3.08}} & {\textbf{3.38}} & {3.82} & {3.62}
& {3.16} & {3.44} & {4.00} & {3.44}
& {3.07} & {3.43} & {3.81} & {3.63}
& {3.11} & {3.47} & {3.82} & {3.71}
& {2.93} & {3.38} & {3.62} & {3.34} \\

& VoiceFixer \cite{liu22voicefixer}
& 2.95 & 3.25 & 3.89 & 3.65
& 2.94 & 3.25 & \textbf{3.83} & 3.55
& {3.13} & {3.43} & {3.98} & {3.46} 
& 2.88 & 3.26 & 3.61 & 3.62
& 3.11 & 3.40 & 3.97 & 3.84
& \multicolumn{4}{c}{-} \\

& {CDiffuse \cite{cdiffuse}}
& {2.22} & {3.22} & {2.28} & {2.70}
& {2.50} & {3.17} & {2.96} & {2.78}
& {2.83} & {3.31} & {3.60} & {2.99}
& {1.62} & {2.03} & {2.07} & {2.63}
& {1.98} & {2.65} & {2.25} & {2.83}
& {1.92} & {2.48} & {2.59} & {2.44} \\

& StoRM \cite{Lemercier2023storm}
& 3.29 & 3.57 & 4.03 & 3.87
& 2.97 & 3.37 & 3.65 & 3.79
& {3.20} & {3.46} & {4.06} & {3.61}
& 3.25 & 3.53 & 4.01 & 4.01
& 3.25 & 3.53 & 4.03 & 3.87
& 3.04 & 3.42 & 3.80 & 3.68 \\

& {UNIVERSE++ \cite{universe++}}
& {3.18} & {3.43} & {4.05} & {3.71}
& {2.97} & {3.32} & {3.73} & {3.63}
& {3.18} & {3.43} & {4.07} & {3.56}
& {3.19} & {3.50} & {3.95} & {3.81}
& {2.74} & {3.34} & {3.28} & {3.54} 
& {2.93} & {3.34} & {3.68} & {3.49} \\

& SpatialNet \cite{quan2024spatialnet}
& 2.76 & 3.32 & 3.39 & 3.51
& 2.86 & 3.27 & 3.58 & 3.59
& {3.02} & {3.33} & {3.92} & {3.53}
& 3.07 & 3.41 & 3.87 & 3.84
& 2.91 & 3.33 & 3.68 & 3.65
& 2.88 & 3.32 & 3.66 & 3.45 \\

& SpatialNet-Mamba
& 2.94 & 3.35 & 3.71 & 3.65
& 2.94 & 3.31 & 3.69 & 3.67
& {3.08} & {3.36} & {3.98} & {3.57}
& 3.09 & 3.43 & 3.85 & 3.86
& 2.89 & 3.35 & 3.58 & 3.68
& 2.95 & 3.37 & 3.69 & 3.50 \\

\cline{2-26}

& CleanMel-S-map
& 3.33 & 3.58 & 4.11 & 3.96
& 3.02 & 3.36 & 3.73 & 3.79
& {\textbf{3.26}} & {3.49} & {\textbf{4.11}} & {\textbf{3.63}}
& \textbf{3.39} & 3.62 & 4.13 & 4.08
& \textbf{3.33} & \textbf{3.58} & 4.10 & 4.03
& \textbf{3.25} & \textbf{3.55} & \textbf{4.03} & \textbf{3.82} \\

& CleanMel-S-mask
& 3.26 & 3.56 & 4.01 & 3.81
& 2.97 & 3.34 & 3.68 & 3.71
& {3.22} & {3.47} & {4.08} & {3.59}
& 3.31 & 3.58 & 4.08 & 4.01
& 3.25 & 3.53 & 4.05 & 3.97
& 3.15 & 3.49 & 3.93 & 3.71 \\

& CleanMel-L-map
& \textbf{3.36} & \textbf{3.60} & \textbf{4.13} & \textbf{3.99}
& 3.02 & {3.37} & 3.73 & \textbf{3.81}
& \textbf{3.26} & \textbf{3.50} & \textbf{4.11} & \textbf{3.63}
& \textbf{3.39} & \textbf{3.63} & \textbf{4.14} & \textbf{4.09}
& \textbf{3.33} & 3.57 & \textbf{4.12} & \textbf{4.05}
& 3.23 & 3.52 & 4.02 & \textbf{3.82} \\

& CleanMel-L-mask
& 3.30 & 3.59 & 4.04 & 3.87
& 2.99 & 3.35 & 3.70 & 3.74
& 3.23 & 3.48 & 4.09 & 3.59
& 3.31 & 3.58 & 4.08 & 4.02
& 3.25 & 3.53 & 4.05 & 4.00
& 3.15 & 3.49 & 3.93 & 3.71 \\

\bottomrule
\end{tabular}

\footnotesize{}
\end{center}
\end{table*}

%% file: FinalFiles/Tables/R_SE_PESQ.tex
\begin{table}[t]
\begin{center}
\setlength\tabcolsep{1pt}
\renewcommand\arraystretch{1.2}
\caption{PESQ scores on test sets with clean reference speech. }
\vspace{-1em}
\label{tab: se_pesq}
\begin{tabular}{c|c|c|c|c|c|c}

\toprule
& 
\multirow{2}{*}{\makecell[t]{\textbf{Enhancement} \\ \textbf{method}}} &
\multirow{2}{*}{\makecell[t]{CHiME4 \\ simu.}} &
\multirow{2}{*}{\makecell[t]{DNS \\ w.o. rvb.}} &
\multirow{2}{*}{\makecell[t]{VoiceBank \\ Demand}} & 
\multirow{2}{*}{\makecell[t]{REVERB\\ simu.}} &
\multirow{2}{*}{\makecell[t]{RealMAN \\ static}} \\

& & & & & \\

\hline
& unproc.
& 1.20 & 1.58 & {1.97} & 1.50 & 1.14 \\
\midrule
\multirow{6}{*}{\rotatebox{90}{\textbf{Online}}}
& FullSubNet \cite{hao2021fullsubnet}
& 1.80 & 2.71 & \textbf{2.73} & 2.39 & 1.54 \\
& Demucs \cite{defossez20demucs}
& 1.64 & 2.63 & {2.55} & 1.92 & - \\
& {LiSenNet \cite{yan2025lisennet}}
& 1.87 & 2.58 & 2.72 & 2.03 & 1.65  \\
& oSpatialNet \cite{quan2024oSpatialNet}
& {1.81} & {2.77} & {2.60} & 2.56 & \textbf{1.87} \\
\cline{2-7}
& CleanMel-S-map
& 1.77 & 2.73 & {2.43} & \textbf{2.63} & 1.79 \\
& CleanMel-S-mask
& \textbf{1.99} & \textbf{2.82} & {2.55} & \textbf{2.63} & 1.75 \\
\midrule
\midrule
\multirow{13}{*}{\rotatebox{90}{\textbf{Offline}}}
& {ConvTasNet} \cite{Luo2019Conv}
& {1.81} & {2.56} & {2.53} & {2.18} & {1.55}  \\
& FullSubNet \cite{hao2021fullsubnet}
& 1.87 & 2.82 & {2.88} & 2.48 & 1.61 \\
& {CMGAN \cite{CMGAN}}
& {1.85} & {2.72} & {2.53} & {2.68} & {1.82}  \\
& VoiceFixer \cite{liu22voicefixer}
& 1.57 & 1.92 & {2.37} & 1.67 & - \\
& {CDiffuse \cite{cdiffuse}}
& {1.54} & {1.90} & {2.39} & {1.42} & {1.13}  \\
& StoRM \cite{Lemercier2023storm}
& 1.76 & 2.74 & {2.73} & 2.52 & 1.71  \\
& {UNIVERSE++ \cite{universe++}}
& {1.89} & {2.74} & {2.87} & {2.42} & {1.70}  \\
& SpatialNet \cite{quan2024spatialnet}
& 1.90 & 2.82 & {2.71} & 2.87 & 1.93 \\
& SpatialNet-Mamba
& 1.90 & 2.90 & {2.72} & \textbf{3.06} & \textbf{2.10} \\
\cline{2-7}
& CleanMel-S-map
& 2.09 & 2.95 & {2.76} & 2.92 & 2.00 \\
& CleanMel-S-mask
& 2.17 & 2.97 & {2.81} & 2.85 & 1.95 \\
& {CleanMel-L-map}
& {2.23} & {3.06} & {2.85} & {3.01} & {\textbf{2.10}} \\
& CleanMel-L-mask
& \textbf{2.35} & \textbf{3.07} & {\textbf{2.89}} & 2.97 & 2.01 \\

\bottomrule
\end{tabular}

\footnotesize{}
\end{center}
\end{table}

%% file: FinalFiles/Figs/R_mask_mapping.tex
\begin{figure}[t]
\begin{center}
    \includegraphics[width=\linewidth]{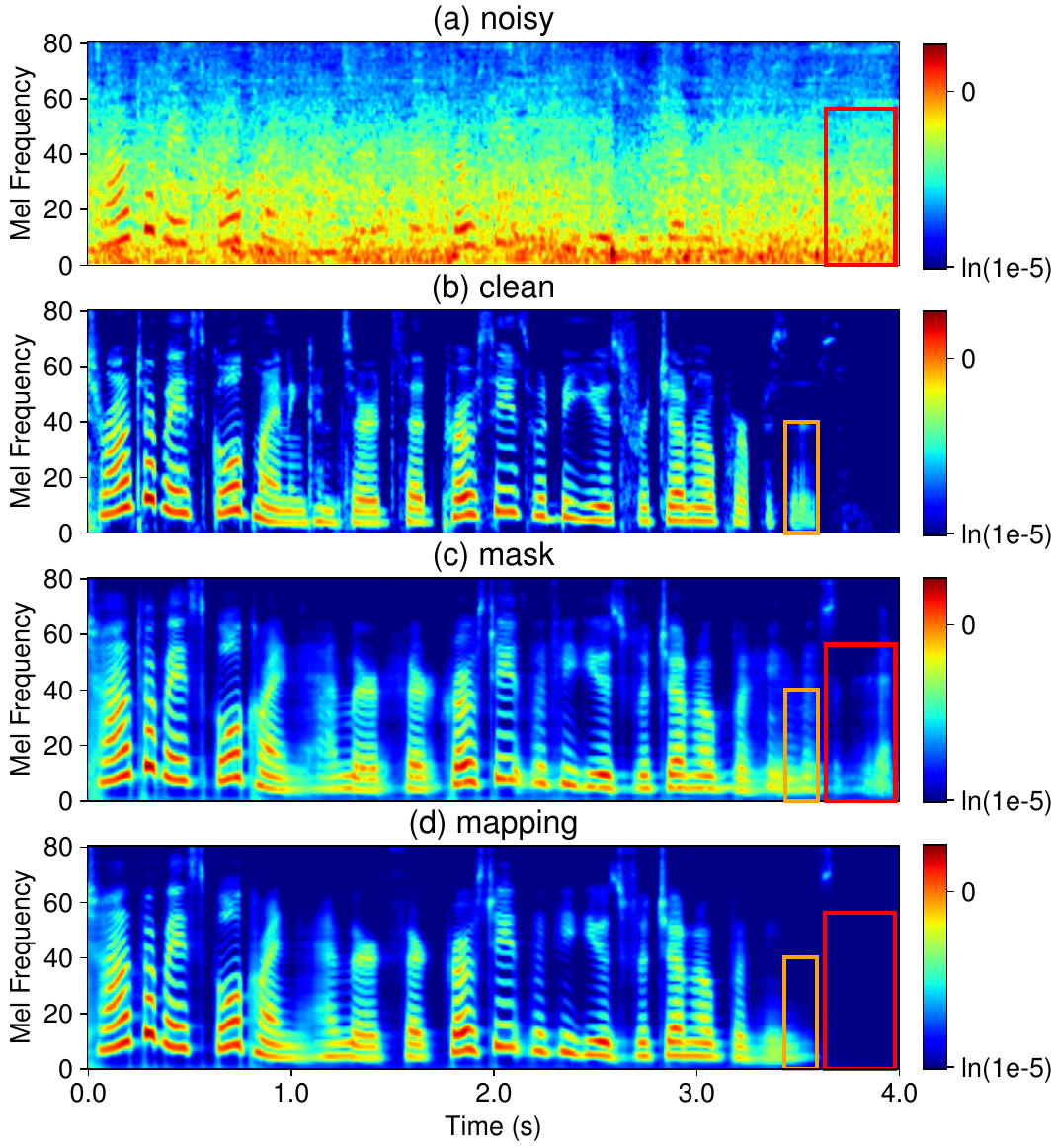}
    \vspace{-2em}
    \caption{An example of Mel spectrogram enhancement with the targets of Mel ratio mask and logMel mapping. The orange box marks the removed part of speech by \emph{mapping} and the red box marks the residual noise by \emph{mask}. }
    
    \label{fig: r_mask_mapping}
\end{center}
\end{figure}

%% file: FinalFiles/Tables/R_ASR.tex
\begin{table}[t]
\begin{center}
\setlength\tabcolsep{2pt}
\renewcommand\arraystretch{1.3}
\caption{ASR results, WER (\%) for the English CHiME4 and REVERB sets, and CER (\%) for the Chinese RealMAN set. }
\vspace{-2em}
\label{tab: asr_all}
\begin{threeparttable}

\begin{tabular}{c|cc|cc|cc|c}

\toprule
\multirow{2}{*}{\makecell[t]{\textbf{Enhancement} \\ \textbf{method}}} &
\multicolumn{2}{c|}{CHiME4} &
\multicolumn{2}{c|}{REVERB simu.} &
\multicolumn{2}{c|}{REVERB real} & 
\multirow{2}{*}{\makecell{RealMAN \\ static}} \\
\cline{2-7}
& simu. & real
& \hspace{0.5em} near & far    
& \hspace{0.5em} near & far
&  \\

\midrule
unprocessed 
& 15.3 & 13.1 & \hspace{0.5em} 3.7 & 4.7 & \hspace{0.5em} 6.0 & 6.3 & 20.1 \\
\hline
clean
& 3.1 & - & \hspace{0.5em} 3.4 & 3.4 & \hspace{0.5em}  - & - & 7.7 \\
\textcolor{gray}{Multi-channel*}
& \textcolor{gray}{13.0} 
& \textcolor{gray}{10.8} 
& \hspace{0.5em}\textcolor{gray}{3.5} 
& \textcolor{gray}{3.7} 
& \hspace{0.5em}\textcolor{gray}{3.6} 
& \textcolor{gray}{4.4} 
& \textcolor{gray}{-}
\\
1ch WPE \cite{Nakatani2015wpe}
& - & - & \hspace{0.5em}3.7 & 4.6 & \hspace{0.5em}5.5 & 5.8 & - \\

{ConvTasNet \cite{Luo2019Conv}}
& {17.8} & {20.8} & \hspace{0.5em}4.6 & 8.8 & \hspace{0.5em}11.4 & 11.9 & {28.7} \\

FullSubNet \cite{hao2021fullsubnet}
& 19.4 & 17.3 & \hspace{0.5em}4.1 & 5.6 & \hspace{0.5em}6.1 & 6.0 & 22.0 \\

{CMGAN \cite{universe++}}
& {18.0} & {14.7} & \hspace{0.5em}{4.1} & {4.5} & \hspace{0.5em}{5.0} & {5.6} & {21.0} \\

VoiceFixer$^{\dag}$ \cite{liu22voicefixer}
& 24.3 & 26.4 & \hspace{0.5em}5.9 & 8.8 & \hspace{0.5em}8.9 & 10.1 & - \\

{CDiffuse \cite{universe++}}
& {18.2} & {17.3} & \hspace{0.5em}{7.8} & {6.8} & \hspace{0.5em}{43.0} & {45.6} & {50.8} \\

StoRM \cite{Lemercier2023storm}
& 21.9 & 18.2 & \hspace{0.5em}4.2 & 5.5 & \hspace{0.5em}7.3 & 7.3 & 28.4\\

{UNIVERSE++ \cite{universe++}}
& {20.8} & {18.6} & \hspace{0.5em}{4.6} & {6.8} & \hspace{0.5em}{8.4} & {9.0} & {32.3} \\

SpatialNet  \cite{quan2024spatialnet} 
& 20.5 & 16.9 & \hspace{0.5em}4.2 & 5.2 & \hspace{0.5em}5.7 & 5.1 & 18.5 \\

SpatialNet-Mamba
& 17.6 & 13.7 & \hspace{0.5em}3.9 & 4.0 & \hspace{0.5em}4.3 & 4.9 & 16.5 \\

\cline{1-8}
CleanMel-S-map
& 12.5 & 9.8 & \hspace{0.5em}3.7 & 3.9 & \hspace{0.5em}4.2 & 4.5 & 16.7 \\
CleanMel-S-mask
& 11.9 & 9.3 & \hspace{0.5em}3.6 & \textbf{3.7} & \hspace{0.5em}3.9 & 4.3 & 16.5 \\
{CleanMel-L-map}
& {10.2} & {8.1} & \hspace{0.5em}{3.7} & {3.8} & \hspace{0.5em}{3.8} & {\textbf{3.3}} & {14.9} \\
CleanMel-L-mask
& \textbf{9.6} & \textbf{7.7} & \hspace{0.5em}\textbf{3.5} & \textbf{3.7} & \hspace{0.5em}\textbf{3.5} & 3.6 & \textbf{14.4} \\

\bottomrule
\end{tabular}

\begin{tablenotes}
\item \textcolor{gray}{* 5-channel Beamformit and 8-channel WPE \cite{Nakatani2015wpe} + Beamformit serve as the multi-channel baseline methods for CHiME4 and REVERB, respectively.}
\item {\dag VoiceFixer is evaluated in the same way as the proposed model, namely the enhanced Mel-spectrogram is directly fed to ASR models. To fit the STFT setups of VoiceFixer, the ASR models used for VoiceFixer were re-trained, which are different from the ones used for other methods. Please refer to Table \ref{tab: asr_wav} for more information about the re-trained ASR models.}
\end{tablenotes}

\end{threeparttable}
\end{center}
\end{table}

%% file: FinalFiles/Figs/R_mel_linear_loss.tex
\begin{figure}[t]
    \centering
    \includegraphics[width=0.9\linewidth]{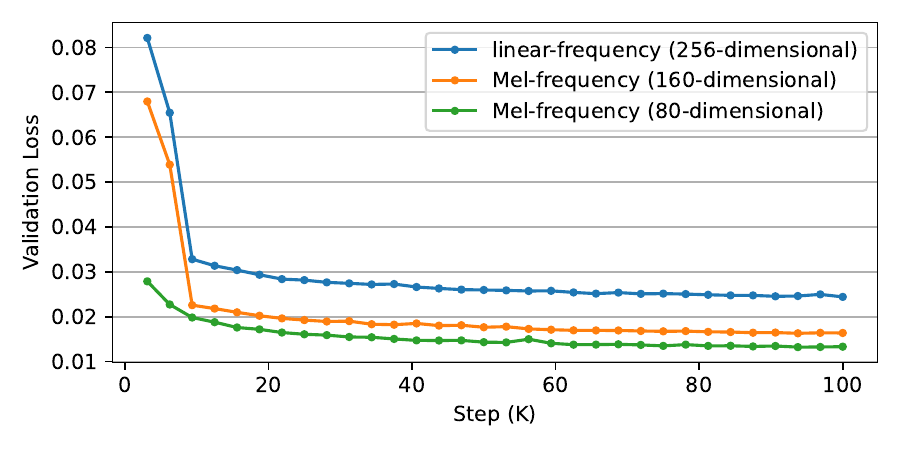}
    \vspace{-1.5em}
    \caption{Comparison of validation losses between linear- and Mel-frequency speech enhancement.}
    \label{fig: mrm_compare}
\end{figure}

%% file: FinalFiles/Tables/R_ASR_clip_value.tex
\begin{table}[t]
\begin{center}
\setlength\tabcolsep{3pt}
\renewcommand\arraystretch{1.3}
\caption{The influence of different logMel clip values ($\epsilon$) to speech quality  (DNSMOS) and ASR (WER).}
\vspace{-1em}
\begin{tabular}{c|cc|cc}
\toprule
\multirow{3}{*}{\makecell[c]{\textbf{Clip values ($\epsilon$)} \\ \textbf{of clean speech}}} & \multicolumn{2}{c|}{CHiME4} & \multicolumn{2}{c}{REVERB} \\
\cline{2-5}
& \multirow{2}{*}{\makecell[c]{DNSMOS\\ P.835 OVL}} & \multirow{2}{*}{\makecell[c]{WER \\ (\%)}} & \multirow{2}{*}{\makecell[c]{DNSMOS\\ P.835 OVL}}  & \multirow{2}{*}{\makecell[c]{WER \\ (\%)}}   \\
& & & & \\
\midrule
$1e-10$ & 3.27 & 3.11 & 3.26  & 3.39   \\
$1e-5$  & 3.27 & 3.16 & 3.28  & 3.39   \\
\bottomrule
\end{tabular}

\label{tab: asr_clip_value}
\end{center}
\end{table}

%% file: FinalFiles/Tables/R_ASR_wav.tex
\begin{table}[t]
\begin{center}
\setlength\tabcolsep{2pt}
\renewcommand\arraystretch{1.2}
\caption{ASR result comparisons between taking as input enhanced Mel-spectrogram and vocoder waveform. WER (\%) for the English CHiME4 and REVERB sets, and CER (\%) for the Chinese RealMAN set.}
\vspace{-1em}
\label{tab: asr_wav}
\begin{tabular}{c|cc|cc|cc|c}

\toprule
\multicolumn{1}{c|}{\multirow{2}{*}{\makecell[t]{\textbf{Enhancement} \\ \textbf{method}}}} &
\multicolumn{2}{c|}{CHiME4} &
\multicolumn{2}{c|}{REVERB simu.} &
\multicolumn{2}{c|}{REVERB real} & 
\multirow{2}{*}{\makecell{RealMAN \\ static}} \\
\cline{2-7}
& simu. & real
& \hspace{0.5em} near & far    
& \hspace{0.5em} near & far
&  \\

\midrule
\multicolumn{1}{c|}{unprocessed}
& 15.6 & 13.5 & \hspace{0.5em}4.3 & 5.6 & \hspace{0.5em}6.8 & 7.4 & - \\
\cline{1-8}
VoiceFixer 
& 24.3 & 26.4 & \hspace{0.5em}5.9 & 8.8 & \hspace{0.5em}8.9 & 10.1 & - \\
waveform
& 27.6 & 30.6 & \hspace{0.5em}8.1 & 11.9 & \hspace{0.5em}10.9 & 11.5 & - \\
\midrule
\midrule
\multicolumn{1}{c|}{unprocessed}
& 15.3 & 13.1 & \hspace{0.5em}3.7 & 4.7 & \hspace{0.5em}6.0 & 6.3 & 20.1 \\
\cline{1-8}
CleanMel-S-mask 
& 11.9 & 9.3 & \hspace{0.5em}3.6 & 3.7 & \hspace{0.5em}3.9 & 4.3 & 16.5 \\
waveform 
& 13.4 & 10.1 & \hspace{0.5em}3.7 & 4.2 & \hspace{0.5em}4.0 & 4.4 & 17.9 \\
CleanMel-L-mask
& 9.6 & 7.7 & \hspace{0.5em}3.5 & 3.7 & \hspace{0.5em}3.5 & 3.6 & 14.4 \\
waveform 
& 10.2 & 8.4 & \hspace{0.5em}3.8 & 3.9 & \hspace{0.5em}3.4 & 3.7 & 15.7 \\

\bottomrule
\end{tabular}

\end{center}
\end{table}

%% file: FinalFiles/Tables/R_ASR_finetune.tex
\begin{table}[t]
\begin{center}
\setlength\tabcolsep{3pt}
\renewcommand\arraystretch{1.3}
\caption{{ASR results (WER, \%) on CHiME4 of fine-tuning CleanMel and/or the ASR model.}}
\vspace{-1em}
\label{tab: asr_ft}
\begin{tabular}{c|cc|cc}
\toprule
\multirow{2}{*}{\makecell[t]{\textbf{Enhancement} \\ \textbf{method}}}
& \multicolumn{2}{c|}{Fine-tuning}
& \multicolumn{2}{c}{CHiME4} \\

\cline{2-5}
& CleanMel & ASR & simu. & real \\

\midrule
unprocessed 
& \multicolumn{2}{c|}{-} & 15.3 & 13.1 \\

\hline

\multirow{4}{*}{CleanMel-S-mask}
& $\times$ & $\times$ & 11.9 & 9.3  \\

& \checkmark & $\times$ & 10.6 & 8.8 \\
& $\times$ & \checkmark & 11.2 & 8.8 \\ 
& \checkmark & \checkmark & 11.1 & 8.8 \\ 

\bottomrule
\end{tabular}

\end{center}
\end{table}

%% file: FinalFiles/Tables/R_model_complexity.tex
\begin{table}[t]
    \centering
    \caption{Model size and computational complexity of different models.}
    \vspace{-1em}
    \label{tab: model_complexity}
    \renewcommand\arraystretch{1.2}
    \begin{tabular}{c|c|c|c}
    \toprule
    & \textbf{Model} & \textbf{\#Param(M)} & \textbf{FLOPs(G/s)} \\
    \midrule
    \multirow{5}{*}{\rotatebox{90}{\textbf{Online}}}
    & FullSubNet \cite{hao2021fullsubnet} & 5.6 & 60 \\
    & Demucs \cite{defossez20demucs} & 33.5 & 15 \\    
    & {LiSenNet} \cite{yan2025lisennet} & {0.04} & {0.1} \\
    & oSpatialNet \cite{quan2024oSpatialNet} & 1.7 & 37 \\
    \cline{2-4}
    & {CleanMel-S + Vocos} & 2.7 + 13.2 & 18 + 2 \\
    \midrule
    \midrule
    \multirow{11}{*}{\rotatebox{90}{\textbf{Offline}}}
    & {ConvTasNet} \cite{Luo2019Conv} & {5.0} & {10} \\
    & FullSubNet \cite{hao2021fullsubnet} & 14.6 & 158 \\
    & {CMGAN} \cite{CMGAN} & {1.8} & {62} \\
    & {VoiceFixer + HifiGAN} \cite{liu22voicefixer} & 88.3 + 33.8 & 21 + 103 \\
    & {CDiffuse} \cite{cdiffuse} & {10.2} & {1520} \\    
    & StoRM \cite{Lemercier2023storm} & 55.1 & 4600 \\
    & {UNIVERSE++} \cite{universe++} & {148.8} & {163} \\
    & SpatialNet \cite{quan2024spatialnet} & 1.6 & 46 \\
    & SpatialNet-Mamba & 1.7 & 35 \\
    \cline{2-4}
    & {CleanMel-S + Vocos} & 2.5 + 13.2 & 33 + 3 \\
    & {CleanMel-L + Vocos} & 7.2 + 13.2 & 128 + 3 \\
    \bottomrule
    \end{tabular}

\end{table}

%% file: FinalFiles/6_conclusion.tex
\section{Conclusion}

This work proposed a single-channel Mel-spectrogram denoising and dereverberation network, named CleanMel. The learning targets of logMel mapping and ratio mask have been compared, while the former suffers from less residual noise and the latter preserves better the target speech. The adopted network architecture, i.e. interleaved cross-band and narrow-band blocks, has been proven working well for single-channel speech denoising and dereverberation in both the linear-frequency domain and the proposed Mel-frequency domain. The high-quality enhanced Mel-spectrogram can be well transformed to waveform with a neural vocoder and can also be used for boosting the ASR performance. Mel-spectrogram plays a key role in the field of speech processing, so the proposed CleanMel model can be potentially used for many other tasks, such as self-supervised speech pre-training and high-quality speech synthesis.   

%% file: main.bbl
\begin{thebibliography}{10}
\providecommand{\url}[1]{#1}
\csname url@samestyle\endcsname
\providecommand{\newblock}{\relax}
\providecommand{\bibinfo}[2]{#2}
\providecommand{\BIBentrySTDinterwordspacing}{\spaceskip=0pt\relax}
\providecommand{\BIBentryALTinterwordstretchfactor}{4}
\providecommand{\BIBentryALTinterwordspacing}{\spaceskip=\fontdimen2\font plus
\BIBentryALTinterwordstretchfactor\fontdimen3\font minus
  \fontdimen4\font\relax}
\providecommand{\BIBforeignlanguage}[2]{{%
\expandafter\ifx\csname l@#1\endcsname\relax
\typeout{** WARNING: IEEEtran.bst: No hyphenation pattern has been}%
\typeout{** loaded for the language `#1'. Using the pattern for}%
\typeout{** the default language instead.}%
\else
\language=\csname l@#1\endcsname
\fi
#2}}
\providecommand{\BIBdecl}{\relax}
\BIBdecl

\bibitem{Luo2019Conv}
Y.~Luo and N.~Mesgarani, ``Conv-tasnet: Surpassing ideal time--frequency
  magnitude masking for speech separation,'' \emph{IEEE/ACM Transactions on
  Audio, Speech, and Language processing}, vol.~27, no.~8, pp. 1256--1266,
  2019.

\bibitem{defossez20demucs}
A.~Défossez, G.~Synnaeve, and Y.~Adi, ``Real time speech enhancement in the
  waveform domain,'' in \emph{Interspeech 2020}, 2020, pp. 3291--3295.

\bibitem{Xiong2022Spectro}
F.~Xiong, W.~Chen, P.~Wang, X.~Li, and J.~Feng, ``Spectro-temporal subnet for
  real-time monaural speech denoising and dereverberation,'' in
  \emph{Interspeech 2022}, 2022, pp. 931--935.

\bibitem{hu2020DCCRN}
Y.~Hu, Y.~Liu, S.~Lv, M.~Xing, S.~Zhang, Y.~Fu, J.~Wu, B.~Zhang, and L.~Xie,
  ``Dccrn: Deep complex convolution recurrent network for phase-aware speech
  enhancement,'' in \emph{Interspeech 2020}, 2020, pp. 2472--2476.

\bibitem{li2022glance}
A.~Li, C.~Zheng, L.~Zhang, and X.~Li, ``Glance and gaze: A collaborative
  learning framework for single-channel speech enhancement,'' \emph{Applied
  Acoustics}, vol. 187, p. 108499, 2022.

\bibitem{iwamoto22analysis}
K.~Iwamoto, T.~Ochiai, M.~Delcroix, R.~Ikeshita, H.~Sato, S.~Araki, and
  S.~Katagiri, ``How bad are artifacts?: Analyzing the impact of speech
  enhancement errors on asr,'' in \emph{Interspeech 2022}, 2022, pp.
  5418--5422.

\bibitem{kinoshita2020improving}
K.~Kinoshita, T.~Ochiai, M.~Delcroix, and T.~Nakatani, ``Improving noise robust
  automatic speech recognition with single-channel time-domain enhancement
  network,'' in \emph{ICASSP 2020 - 2020 IEEE International Conference on
  Acoustics, Speech and Signal Processing (ICASSP)}, 2020, pp. 7009--7013.

\bibitem{nian2022progressive}
Z.~Nian, J.~Du, Y.~Ting~Yeung, and R.~Wang, ``A time domain progressive
  learning approach with snr constriction for single-channel speech enhancement
  and recognition,'' in \emph{ICASSP 2022 - 2022 IEEE International Conference
  on Acoustics, Speech and Signal Processing (ICASSP)}, 2022, pp. 6277--6281.

\bibitem{yang2023robustASR}
Y.~Yang, A.~Pandey, and D.~Wang, ``Time-domain speech enhancement for robust
  automatic speech recognition,'' in \emph{Interspeech 2023}, 2023, pp.
  4913--4917.

\bibitem{yang2024towards}
------, ``Towards decoupling frontend enhancement and backend recognition in
  monaural robust asr,'' \emph{arXiv preprint arXiv:2403.06387}, 2024.

\bibitem{valin2018hybrid}
J.-M. Valin, ``A hybrid dsp/deep learning approach to real-time full-band
  speech enhancement,'' in \emph{2018 IEEE 20th International Workshop on
  Multimedia Signal Processing (MMSP)}, 2018, pp. 1--5.

\bibitem{valin2020perceptually}
J.-M. Valin, U.~Isik, N.~Phansalkar, R.~Giri, K.~Helwani, and A.~Krishnaswamy,
  ``A perceptually-motivated approach for low-complexity, real-time enhancement
  of fullband speech,'' in \emph{Interspeech 2020}, 2020, pp. 2482--2486.

\bibitem{schroter23deepfilter}
H.~Schröter, A.~N. Escalante-B., T.~Rosenkranz, and A.~Maier, ``Deepfilternet:
  Perceptually motivated real-time speech enhancement,'' in \emph{Interspeech
  2023}, 2023, pp. 2008--2009.

\bibitem{li2019narrow}
X.~Li and R.~Horaud, ``Narrow-band deep filtering for multichannel speech
  enhancement,'' \emph{arXiv preprint arXiv:1911.10791}, 2019.

\bibitem{hao2021fullsubnet}
X.~Hao, X.~Su, R.~Horaud, and X.~Li, ``Fullsubnet: A full-band and sub-band
  fusion model for real-time single-channel speech enhancement,'' in
  \emph{ICASSP 2021 - 2021 IEEE International Conference on Acoustics, Speech
  and Signal Processing (ICASSP)}, 2021, pp. 6633--6637.

\bibitem{zhou2023rts}
R.~Zhou, W.~Zhu, and X.~Li, ``Speech dereverberation with a reverberation time
  shortening target,'' in \emph{ICASSP 2023 - 2023 IEEE International
  Conference on Acoustics, Speech and Signal Processing (ICASSP)}, 2023, pp.
  1--5.

\bibitem{hao2022fast}
X.~Hao and X.~Li, ``Fast fullsubnet: Accelerate full-band and sub-band fusion
  model for single-channel speech enhancement,'' \emph{arXiv preprint
  arXiv:2212.09019}, 2022.

\bibitem{kothapally2023deep}
V.~Kothapally, Y.~Xu, M.~Yu, S.-X. Zhang, and D.~Yu, ``Deep neural mel-subband
  beamformer for in-car speech separation,'' in \emph{ICASSP 2023 - 2023 IEEE
  International Conference on Acoustics, Speech and Signal Processing
  (ICASSP)}, 2023, pp. 1--5.

\bibitem{liu22voicefixer}
H.~Liu, X.~Liu, Q.~Kong, Q.~Tian, Y.~Zhao, D.~Wang, C.~Huang, and Y.~Wang,
  ``Voicefixer: A unified framework for high-fidelity speech restoration,'' in
  \emph{Interspeech 2022}, 2022, pp. 4232--4236.

\bibitem{tian2023diffusion}
Y.~Tian, W.~Liu, and T.~Lee, ``Diffusion-based mel-spectrogram enhancement for
  personalized speech synthesis with found data,'' in \emph{2023 IEEE Automatic
  Speech Recognition and Understanding Workshop (ASRU)}, 2023, pp. 1--7.

\bibitem{quan2024spatialnet}
C.~Quan and X.~Li, ``Spatialnet: Extensively learning spatial information for
  multichannel joint speech separation, denoising and dereverberation,''
  \emph{IEEE/ACM Transactions on Audio, Speech, and Language Processing},
  vol.~32, pp. 1310--1323, 2024.

\bibitem{quan2024oSpatialNet}
------, ``Multichannel long-term streaming neural speech enhancement for static
  and moving speakers,'' \emph{IEEE Signal Processing Letters}, vol.~31, pp.
  2295--2299, 2024.

\bibitem{li2019multichannel}
X.~Li, L.~Girin, S.~Gannot, and R.~Horaud, ``Multichannel speech separation and
  enhancement using the convolutive transfer function,'' \emph{IEEE/ACM
  Transactions on Audio, Speech, and Language Processing}, vol.~27, no.~3, pp.
  645--659, 2019.

\bibitem{wang2018supervised}
D.~Wang and J.~Chen, ``Supervised speech separation based on deep learning: An
  overview,'' \emph{IEEE/ACM Transactions on Audio, Speech, and Language
  Processing}, vol.~26, no.~10, pp. 1702--1726, 2018.

\bibitem{gu2023mamba}
A.~Gu and T.~Dao, ``Mamba: Linear-time sequence modeling with selective state
  spaces,'' in \emph{First Conference on Language Modeling}, 2024.

\bibitem{siuzdak2024vocos}
H.~Siuzdak, ``Vocos: Closing the gap between time-domain and fourier-based
  neural vocoders for high-quality audio synthesis,'' in \emph{The Twelfth
  International Conference on Learning Representations}, 2024.

\bibitem{kong2020hifi}
J.~Kong, J.~Kim, and J.~Bae, ``Hifi-gan: generative adversarial networks for
  efficient and high fidelity speech synthesis,'' in \emph{Proceedings of the
  34th International Conference on Neural Information Processing Systems},
  2020.

\bibitem{bu2017aishell1}
H.~Bu, J.~Du, X.~Na, B.~Wu, and H.~Zheng, ``Aishell-1: An open-source mandarin
  speech corpus and a speech recognition baseline,'' in \emph{2017 20th
  conference of the oriental chapter of the international coordinating
  committee on speech databases and speech I/O systems and assessment
  (O-COCOSDA)}.\hskip 1em plus 0.5em minus 0.4em\relax IEEE, 2017, pp. 1--5.

\bibitem{du2017aishell2}
J.~{Du}, X.~{Na}, X.~{Liu}, and H.~{Bu}, ``{AISHELL-2: Transforming Mandarin
  ASR Research Into Industrial Scale},'' \emph{ArXiv}, Aug. 2018.

\bibitem{wang2015thchs}
D.~Wang and X.~Zhang, ``Thchs-30: A free chinese speech corpus,'' \emph{arXiv
  preprint arXiv:1512.01882}, 2015.

\bibitem{richter2024ears}
J.~Richter, Y.-C. Wu, S.~Krenn, S.~Welker, B.~Lay, S.~Watanabe, A.~Richard, and
  T.~Gerkmann, ``Ears: An anechoic fullband speech dataset benchmarked for
  speech enhancement and dereverberation,'' in \emph{Interspeech 2024}, 2024,
  pp. 4873--4877.

\bibitem{voicebank}
C.~Veaux, J.~Yamagishi, and S.~King, ``The voice bank corpus: Design,
  collection and data analysis of a large regional accent speech database,'' in
  \emph{2013 International Conference Oriental COCOSDA held jointly with 2013
  Conference on Asian Spoken Language Research and Evaluation
  (O-COCOSDA/CASLRE)}, 2013, pp. 1--4.

\bibitem{reddy2020dns1}
C.~K. Reddy, V.~Gopal, R.~Cutler, E.~Beyrami, R.~Cheng, H.~Dubey,
  S.~Matusevych, R.~Aichner, A.~Aazami, S.~Braun, P.~Rana, S.~Srinivasan, and
  J.~Gehrke, ``The interspeech 2020 deep noise suppression challenge: Datasets,
  subjective testing framework, and challenge results,'' in \emph{Interspeech
  2020}, 2020, pp. 2492--2496.

\bibitem{reddy2022dnsmos835}
C.~K. Reddy, V.~Gopal, and R.~Cutler, ``Dnsmos p. 835: A non-intrusive
  perceptual objective speech quality metric to evaluate noise suppressors,''
  in \emph{ICASSP 2022-2022 IEEE International Conference on Acoustics, Speech
  and Signal Processing (ICASSP)}.\hskip 1em plus 0.5em minus 0.4em\relax IEEE,
  2022, pp. 886--890.

\bibitem{vctk-demand}
J.~Thiemann, N.~Ito, and E.~Vincent, ``The diverse environments multi-channel
  acoustic noise database (demand): A database of multichannel environmental
  noise recordings,'' in \emph{Proceedings of Meetings on Acoustics}, vol.~19,
  no.~1.\hskip 1em plus 0.5em minus 0.4em\relax AIP Publishing, 2013.

\bibitem{eaton2015ace}
J.~Eaton, N.~D. Gaubitch, A.~H. Moore, and P.~A. Naylor, ``Estimation of room
  acoustic parameters: The ace challenge,'' \emph{IEEE/ACM Transactions on
  Audio, Speech, and Language Processing}, vol.~24, no.~10, pp. 1681--1693,
  2016.

\bibitem{jeub2009air}
M.~Jeub, M.~Schafer, and P.~Vary, ``A binaural room impulse response database
  for the evaluation of dereverberation algorithms,'' in \emph{2009 16th
  International Conference on Digital Signal Processing}, 2009, pp. 1--5.

\bibitem{prawda2022ARNI}
K.~Prawda, S.~J. Schlecht, and V.~V{\"a}lim{\"a}ki, ``Calibrating the sabine
  and eyring formulas,'' \emph{The Journal of the Acoustical Society of
  America}, vol. 152, no.~2, pp. 1158--1169, 2022.

\bibitem{carlo2021dechorate}
D.~D. Carlo, P.~Tandeitnik, C.~Foy, N.~Bertin, A.~Deleforge, and S.~Gannot,
  ``dechorate: a calibrated room impulse response dataset for echo-aware signal
  processing,'' \emph{EURASIP Journal on Audio, Speech, and Music Processing},
  vol. 2021, pp. 1--15, 2021.

\bibitem{Hadad2014MultiChannel}
E.~Hadad, F.~Heese, P.~Vary, and S.~Gannot, ``Multichannel audio database in
  various acoustic environments,'' in \emph{2014 14th International Workshop on
  Acoustic Signal Enhancement (IWAENC)}, 2014, pp. 313--317.

\bibitem{james2016naturalreverb}
J.~Traer and J.~H. McDermott, ``Statistics of natural reverberation enable
  perceptual separation of sound and space,'' \emph{Proceedings of the National
  Academy of Sciences}, vol. 113, no.~48, pp. E7856--E7865, 2016.

\bibitem{Kinoshita2013REVERB}
K.~Kinoshita, M.~Delcroix, T.~Yoshioka, T.~Nakatani, E.~Habets, R.~Haeb-Umbach,
  V.~Leutnant, A.~Sehr, W.~Kellermann, R.~Maas, S.~Gannot, and B.~Raj, ``The
  reverb challenge: A common evaluation framework for dereverberation and
  recognition of reverberant speech,'' in \emph{2013 IEEE Workshop on
  Applications of Signal Processing to Audio and Acoustics}, 2013, pp. 1--4.

\bibitem{nakamura2000rwcp}
S.~Nakamura, ``Acoustic sound database collected for hands-free speech
  recognition and sound scene understanding,'' in \emph{International Workshop
  on Hands-Free Speech Communication}, 2001, pp. 43--46.

\bibitem{reddy2021dns3}
C.~K. Reddy, H.~Dubey, K.~Koishida, A.~Nair, V.~Gopal, R.~Cutler, S.~Braun,
  H.~Gamper, R.~Aichner, and S.~Srinivasan, ``Interspeech 2021 deep noise
  suppression challenge,'' in \emph{Interspeech 2021}, 2021, pp. 2796--2800.

\bibitem{Bing2024RealMAN}
B.~Yang, C.~Quan, Y.~Wang, P.~Wang, Y.~Yang, Y.~Fang, N.~Shao, H.~Bu, X.~Xu,
  and X.~Li, ``Real{MAN}: A real-recorded and annotated microphone array
  dataset for dynamic speech enhancement and localization,'' in \emph{The
  Thirty-eight Conference on Neural Information Processing Systems Datasets and
  Benchmarks Track}, 2024.

\bibitem{audioset}
J.~F. Gemmeke, D.~P.~W. Ellis, D.~Freedman, A.~Jansen, W.~Lawrence, R.~C.
  Moore, M.~Plakal, and M.~Ritter, ``Audio set: An ontology and human-labeled
  dataset for audio events,'' in \emph{ICASSP 2017 - 2017 IEEE International
  Conference on Acoustics, Speech and Signal Processing (ICASSP)}, 2017, pp.
  776--780.

\bibitem{vincent2016chime4}
E.~Vincent, S.~Watanabe, J.~Barker, and R.~Marxer, ``The 4th chime speech
  separation and recognition challenge,'' \emph{URL: http://spandh. dcs. shef.
  ac. uk/chime\_challenge/(last accessed on 1 August, 2018)}, 2016.

\bibitem{Zhang2022wenet}
B.~Zhang, H.~Lv, P.~Guo, Q.~Shao, C.~Yang, L.~Xie, X.~Xu, H.~Bu, X.~Chen,
  C.~Zeng, D.~Wu, and Z.~Peng, ``Wenetspeech: A 10000+ hours multi-domain
  mandarin corpus for speech recognition,'' in \emph{ICASSP 2022 - 2022 IEEE
  International Conference on Acoustics, Speech and Signal Processing
  (ICASSP)}, 2022, pp. 6182--6186.

\bibitem{kim2023branchformer}
K.~Kim, F.~Wu, Y.~Peng, J.~Pan, P.~Sridhar, K.~J. Han, and S.~Watanabe,
  ``E-branchformer: Branchformer with enhanced merging for speech
  recognition,'' in \emph{2022 IEEE Spoken Language Technology Workshop
  (SLT)}.\hskip 1em plus 0.5em minus 0.4em\relax IEEE, 2023, pp. 84--91.

\bibitem{vaswani2017attention}
A.~Vaswani, N.~Shazeer, N.~Parmar, J.~Uszkoreit, L.~Jones, A.~N. Gomez,
  {\L}.~Kaiser, and I.~Polosukhin, ``Attention is all you need,''
  \emph{Advances in neural information processing systems}, vol.~30, 2017.

\bibitem{loshchilov2018AdamW}
I.~Loshchilov and F.~Hutter, ``Decoupled weight decay regularization,'' in
  \emph{International Conference on Learning Representations}, 2019.

\bibitem{rix2001pesq}
A.~W. Rix, J.~G. Beerends, M.~P. Hollier, and A.~P. Hekstra, ``Perceptual
  evaluation of speech quality (pesq)-a new method for speech quality
  assessment of telephone networks and codecs,'' in \emph{ICASSP 2001 - 2001
  IEEE international conference on acoustics, speech, and signal processing
  (ICASSP)}, vol.~2.\hskip 1em plus 0.5em minus 0.4em\relax IEEE, 2001, pp.
  749--752.

\bibitem{reddy2021dnsmos808}
C.~K. Reddy, V.~Gopal, and R.~Cutler, ``Dnsmos: A non-intrusive perceptual
  objective speech quality metric to evaluate noise suppressors,'' in
  \emph{ICASSP 2021-2021 IEEE International Conference on Acoustics, Speech and
  Signal Processing (ICASSP)}.\hskip 1em plus 0.5em minus 0.4em\relax IEEE,
  2021, pp. 6493--6497.

\bibitem{CMGAN}
S.~Abdulatif, R.~Cao, and B.~Yang, ``Cmgan: Conformer-based metric-gan for
  monaural speech enhancement,'' \emph{IEEE/ACM Transactions on Audio, Speech,
  and Language Processing}, vol.~32, pp. 2477--2493, 2024.

\bibitem{yan2025lisennet}
H.~Yan, J.~Zhang, C.~Fan, Y.~Zhou, and P.~Liu, ``Lisennet: Lightweight sub-band
  and dual-path modeling for real-time speech enhancement,'' in \emph{ICASSP
  2025-2025 IEEE International Conference on Acoustics, Speech and Signal
  Processing (ICASSP)}.\hskip 1em plus 0.5em minus 0.4em\relax IEEE, 2025, pp.
  1--5.

\bibitem{cdiffuse}
Y.-J. Lu, Z.-Q. Wang, S.~Watanabe, A.~Richard, C.~Yu, and Y.~Tsao,
  ``Conditional diffusion probabilistic model for speech enhancement,'' in
  \emph{ICASSP 2022-2022 IEEE International Conference on Acoustics, Speech and
  Signal Processing (ICASSP)}.\hskip 1em plus 0.5em minus 0.4em\relax Ieee,
  2022, pp. 7402--7406.

\bibitem{Lemercier2023storm}
J.-M. Lemercier, J.~Richter, S.~Welker, and T.~Gerkmann, ``Storm: A
  diffusion-based stochastic regeneration model for speech enhancement and
  dereverberation,'' \emph{IEEE/ACM Transactions on Audio, Speech, and Language
  Processing}, vol.~31, pp. 2724--2737, 2023.

\bibitem{universe++}
R.~Scheibler, Y.~Fujita, Y.~Shirahata, and T.~Komatsu, ``Universal score-based
  speech enhancement with high content preservation,'' in \emph{Interspeech
  2024}, 2024, pp. 1165--1169.

\bibitem{Nakatani2015wpe}
T.~Nakatani, T.~Yoshioka, K.~Kinoshita, M.~Miyoshi, and B.-H. Juang, ``Speech
  dereverberation based on variance-normalized delayed linear prediction,''
  \emph{IEEE Transactions on Audio, Speech, and Language Processing}, vol.~18,
  no.~7, pp. 1717--1731, 2010.

\bibitem{chime_clean}
J.~S. Garofolo, D.~Graff, D.~Paul, and D.~Pallett, ``Csr-i (wsj0) complete,''
  \emph{(No Title)}, 2007.

\bibitem{reverb_clean}
T.~Robinson, J.~Fransen, D.~Pye, J.~Foote, and S.~Renals, ``Wsjcamo: a british
  english speech corpus for large vocabulary continuous speech recognition,''
  in \emph{1995 International Conference on Acoustics, Speech, and Signal
  Processing}, vol.~1.\hskip 1em plus 0.5em minus 0.4em\relax IEEE, 1995, pp.
  81--84.

\bibitem{chang2022end}
X.~Chang, T.~Maekaku, Y.~Fujita, and S.~Watanabe, ``End-to-end integration of
  speech recognition, speech enhancement, and self-supervised learning
  representation,'' \emph{Interspeech 2022}, 2022.

\bibitem{shi2024waveform}
H.~Shi, M.~Mimura, and T.~Kawahara, ``Waveform-domain speech enhancement using
  spectrogram encoding for robust speech recognition,'' \emph{IEEE/ACM
  Transactions on Audio, Speech, and Language Processing}, 2024.

\bibitem{park19specaug}
D.~S. Park, W.~Chan, Y.~Zhang, C.-C. Chiu, B.~Zoph, E.~D. Cubuk, and Q.~V. Le,
  ``Specaugment: A simple data augmentation method for automatic speech
  recognition,'' in \emph{Interspeech 2019}, 2019, pp. 2613--2617.

\end{thebibliography}
